\documentclass[aps,prl,final,twocolumn,letterpaper]{revtex4}

\usepackage{graphicx}   
\usepackage{import}                         
\usepackage{epstopdf}
\usepackage{amsmath} 
\usepackage{float} 
\usepackage{bm}
\usepackage{amssymb}
\usepackage{quotes}
\usepackage{indentfirst}
\usepackage{color}
\usepackage{transparent}
\usepackage{dcolumn}
\usepackage{braket}
\usepackage{multirow}
\usepackage{cancel} 
\usepackage{mdframed}
\usepackage{color}
\usepackage{bm}
\usepackage{dsfont}
\usepackage{slashed}
\usepackage{soul, color} 
\soulregister\ref{7}  
\soulregister\cite{7} 
\renewcommand{\st}[1]{}

\usepackage{xr}
\usepackage{mathtools}

\usepackage{natbib}

\begin{document}
\rmfamily

\title{Strong coupling and single-photon nonlinearity in free-electron quantum optics}

\author{Aviv~Karnieli$^{1,\ddagger}$}
\email{karnieli@stanford.edu}
\author{Charles~Roques-Carmes$^{1,\ddagger}$}
\email{chrc@stanford.edu}
\author{Nicholas~Rivera$^{2}$}
\author{Shanhui~Fan$^{1}$}

\affiliation{$^{1}$ E. L. Ginzton Laboratories, Stanford University, 348 Via Pueblo, Stanford, CA USA}
\affiliation{$^{2}$ Department of Physics, Harvard University, Cambridge, MA 02138, USA}
\affiliation{$^{\ddagger}$ These authors contributed equally to the work.}

\begin{abstract}
The observation that free electrons can interact coherently with quantized electromagnetic fields and matter systems has led to a plethora of proposals leveraging the unique quantum properties of free electrons. 
At the heart of these proposals lies the assumption of a strong quantum interaction between a flying free electron and a photonic mode. However, existing schemes are intrinsically limited by electron diffraction, which puts an upper bound on the interaction length and therefore the quantum coupling strength. Here, we propose the use of "free-electron fibers'': effectively one-dimensional photonic systems where free electrons co-propagate with two guided modes. The first mode applies a ponderomotive trap to the free electron, effectively lifting the limitations due to electron diffraction. The second mode strongly couples to the guided free electron, with an enhanced coupling that is orders of magnitude larger than previous designs. Moreover, the extended interaction lengths enabled by our scheme allows for strong single-photon nonlinearities mediated by free electrons. We predict a few interesting observable quantum effects in our system, such as deterministic single-photon emission and complex, nonlinear multimode dynamics. Our proposal paves the way towards the realization of many anticipated effects in free-electron quantum optics, such as non-Gaussian light generation, deterministic single photon emission, and quantum gates controlled by free-electron--photon interactions.
\end{abstract}


\maketitle

\section{Introduction}
The study of interactions between flying free electrons and quantized systems, such as  electromagnetic fields or bound electrons, has lead to the emerging field of \textit{free-electron quantum optics} \cite{Kfir2019EntanglementsRegime,DiGiulio2019ProbingElectrons,Gover2020Free-Electron-Bound-ElectronInteraction,BenHayun2021ShapingElectrons,Ruimy2021TowardElectrons,Zhao2021QuantumInteraction,Dahan2021ImprintingElectrons,Feist2022Cavity-mediatedPairs}. 
Free-electron quantum optics provides a promising platform for the generation of nonclassical light~\cite{Kfir2019EntanglementsRegime,DiGiulio2019ProbingElectrons,BenHayun2021ShapingElectrons,Dahan2023CreationElectrons, Adiv2023ObservationRadiation, Feist2022Cavity-mediatedPairs, Karnieli2023Jaynes-CummingsPhotons}, quantum information processing \cite{Reinhardt2021Free-ElectronQubits, Tsarev2021Free-electronRevivals, Baranes2022FreePhotons, Karnieli2024UniversalBlockade, Baranes2023Free-electronCorrection}, and quantum sensing \cite{DiGiulio2019ProbingElectrons, Gorlach2020UltrafastElectrons, Gover2020Free-Electron-Bound-ElectronInteraction, Ruimy2021TowardElectrons, Zhao2021QuantumInteraction, Karnieli2023QuantumElectrons, Bucher2023CoherentlyInterferometer, Bucher2023Free-ElectronNearfields}. Specifically, measurement of free electrons' energy can be used for the heralded generation of quantum light, such as Fock states \cite{Bendana2011Single-photonBeams,Kfir2019EntanglementsRegime,BenHayun2021ShapingElectrons,Adiv2023ObservationRadiation, Feist2022Cavity-mediatedPairs}, squeezed states \cite{DiGiulio2022Optical-cavityElectrons} and even cat and GKP states \cite{Dahan2023CreationElectrons}. More recently, it was predicted that free electrons could serve as ancillary qubits for quantum computation \cite{Karnieli2024UniversalBlockade,Baranes2023Free-electronCorrection}, and provide a platform for nonlinear electron dynamics \cite{Talebi2020StrongPrinciples,Eldar2024Self-TrappingDomain} and deterministic single-photon nonlinearities ~\cite{Karnieli2023Jaynes-CummingsPhotons, Pan2023Low-energyApplications, GarciaDeAbajo2022CompleteElectrons, Synanidis2024QuantumLight} owing to quantum recoil \cite{Tsesses2017,Huang2023QuantumLattices}. 

Such free-electron-induced nonlinearities could become a potential resource for quantum nonlinear optics~\cite{Chang2014QuantumPhoton}. Unlike bound-electron systems, the vast spectral range spanned by free electron emitters (from THz \cite{Korbly2005ObservationRadiation} to x-ray \cite{Shentcis2020TunableMaterials}), their phase-matching bandwidth allowing robustness to cavity detunings \cite{Karnieli2024UniversalBlockade}, as well as their ability to image systems with femtosecond temporal resolution and sub-nanometer spatial resolution \cite{Polman2019,GarciadeAbajo2021OpticalChallengesandOpportunities,Roques-Carmes2023Free-electronlightNanophotonics}, may allow new paradigms in quantum optical technologies.

\begin{figure} [h!]
    \centering
    \includegraphics[scale = 0.7]{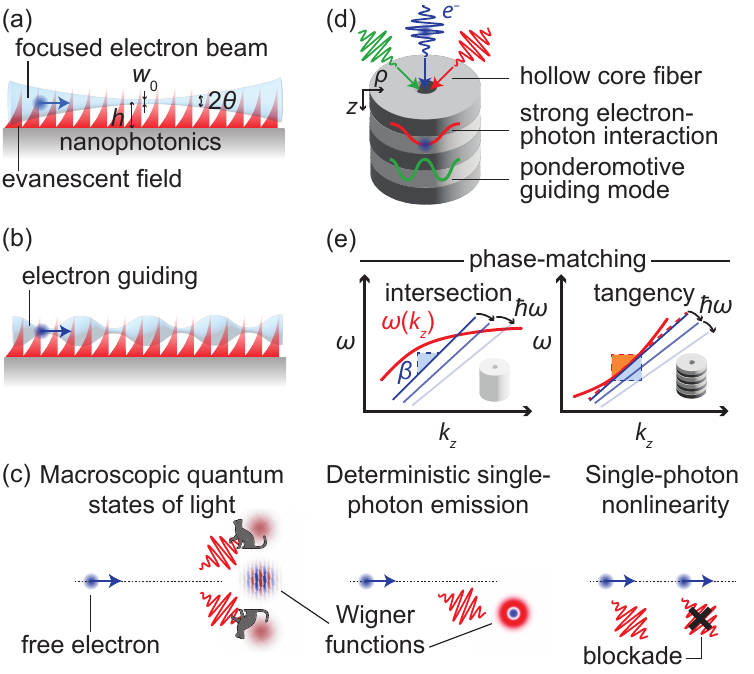}
    \caption{\textbf{Enhanced free-electron-light interactions via ponderomotive guiding.} (a) In conventional free-electron-light interactions, the electron diffraction sets an upper bound on the electron mean free path, and consequently, on the interaction strength. (b) Using ponderomotive electron guiding, one can increase the electron mean free path by counteracting diffraction, resulting in enhanced interaction strengths. (c) Applications of strong interactions between free electrons and light. Representative Wigner functions and input-output pulses are shown for illustration. (d) Schematic of free-electron interactions with modes of an optical fiber providing guiding and strong free-electron--photon interaction. (e) Two schemes for phase-matching: intersection phase-matching in uniform fibers (left) and tangential phase-matching in Bragg fibers (right). The faded lines show the influence of photon emission on detuning.}
    \label{fig:concept}
\end{figure}

In order to realize the above-mentioned advancements, one needs strong free-electron--photon coupling and, sometimes, strong free-electron induced nonlinearities, ideally in a lossless dielectric structure. Free-electron--photon coupling is quantified by the dimensionless parameter $g_Q$, where $|g_Q|^2$ dictates the total number of photons that can be emitted by an electron phase-matched to electromagnetic field modes. The condition $|g_Q|>1$ is considered as the strong coupling regime. Recent experiments reported values of $|g_Q|\approx1$ in electron-plasmon interactions \cite{Adiv2023ObservationRadiation} and $|g_Q|\approx0.1$ in dielectric resonators \cite{Feist2022Cavity-mediatedPairs,Bezard2023HighCavities}, while theoretical proposals have further considered waveguide geometries \cite{Huang2023Electron-PhotonCircuits,DMello2023EfficientWaveguide} reaching similar values. Free-electron-induced nonlinearity, on the other hand, results from the fact that an electron's velocity changes as it absorbs or emits photons \cite{Tsesses2017, Huang2023QuantumLattices}, an effect which can suppress or enhance subsequent photon emission by the same electron, due to modification of energy-momentum conservation \cite{Karnieli2023Jaynes-CummingsPhotons,Pan2023Low-energyApplications}, the so-called \textit{quantum recoil} correction \cite{Tsesses2017}. The nonlinearity is characterized by a nonlinear phase $\delta_{\mathrm{NL}}$, where the condition for strong nonlinearity is $\delta_{\mathrm{NL}} \sim 2\pi$.  The quantum recoil correction has only been recently observed in the x-ray \cite{Huang2023QuantumLattices}, while in the visible-IR range, such free-electron-induced nonlinearity has only been considered theoretically for very low energy electrons (e.g.,$\leq~1\mathrm{keV}$ kinetic energy) \cite{Karnieli2023Jaynes-CummingsPhotons,Pan2023Low-energyApplications}, which introduces many additional experimental challenges which are yet to be overcome.

The ultimate factor limiting both the coupling strength $|g_Q|$ and the nonlinear phase $\delta_{\mathrm{NL}}$ is the \textit{finite interaction length}, $L_{\mathrm{int}}$, between the free electron and the photon. While there are several practical factors hampering the interaction length in experiments, such as beam-sample alignment and sample charging, there exists a fundamental limitation due to electron beam diffraction, which causes the electron to eventually collide with the structure (after such collision, the electron is absorbed in the material and thereby lost, whereas incoherent light emission through electron-matter interaction can become dominant~\cite{Brenny2014QuantifyingMetals}). From simplified geometric arguments (Fig.~\ref{fig:concept}(a)), the interaction length is proportional to $2h/\theta$, where $h$ is the beam position above the structure and $\theta$ its divergence angle, which vanishes as the electron velocity decreases (see Appendix A). Unfortunately, the latter observation can seriously hamper the realization of free-electron-induced nonlinearity effects: especially in experiments with low electron energies \cite{Massuda2018Smith-PurcellElectrons,Roitman2024CoherentBeams}. Thus, venturing into the regimes of strong coupling and single-photon nonlinearity requires a significant increase in interaction lengths across a vast range of electron energies.

\begin{figure*}
    \centering
    \includegraphics[scale = 0.75]{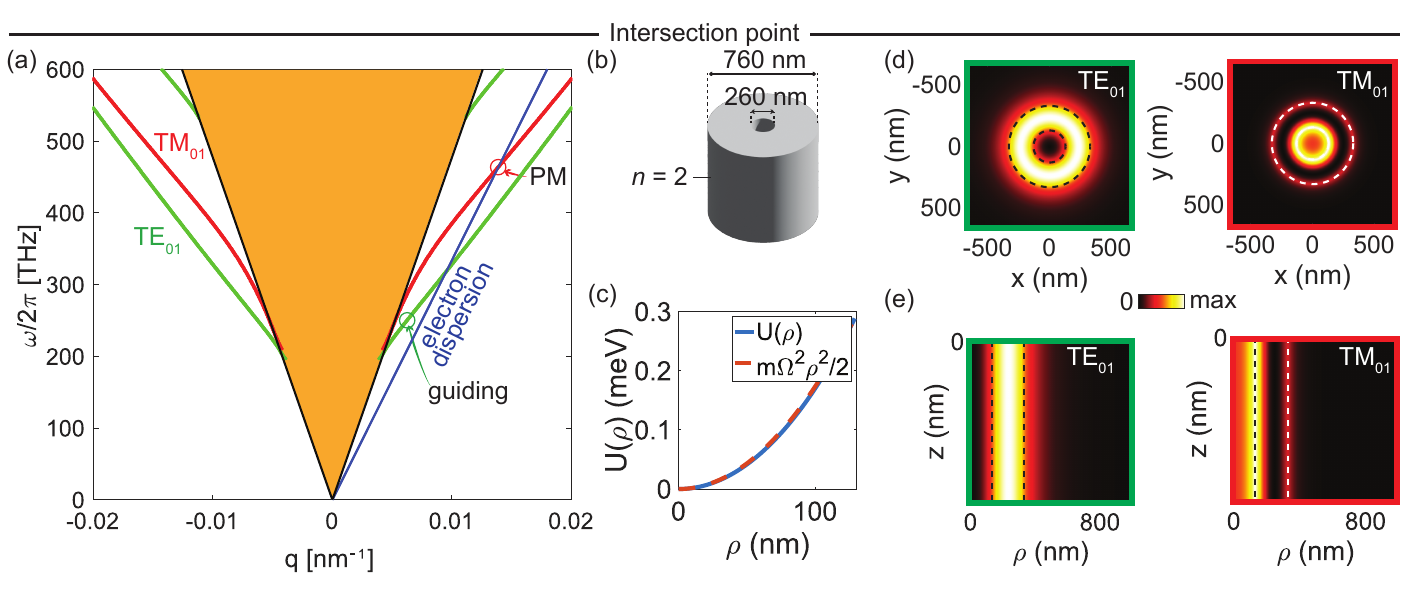}
    \caption{\textbf{Intersection phase-matching in free-electron ponderomotive guiding.} (a) TE and TM band structure of a uniform hollow-core nanofiber design. (b) Schematic of the corresponding design. (c) Corresponding ponderomotive potential from the TE mode with peak pump power $P_0 = 30~\text{W}$. (d) Transverse cross sections of mode profiles for TE ($|E_\varphi|$) and TM ($|E_z|$) modes. (e) Longitudinal cross sections of mode profiles for TE ($|E_\varphi|$) and TM ($|E_z|$) modes. In (d) and (e), dashed lines indicate inner and outer fiber boundaries. PM: phase matching. Parameters used in the calculation are $n=2$ (achievable with tellurite glass~\cite{Yang2006PhotonicGlasses, Liao2012DirectlyLength, Ohishi2012FlatFiber}); $a=180~\mathrm{nm}$, $b=380~\mathrm{nm}$, and electron energy $E=200 \mathrm{keV}$.}
    \label{fig:bandstructures-intersection}
\end{figure*}

In this work, we propose a general method to achieve both strong free-electron--photon coupling and large free-electron-induced nonlinearity on the single-photon level in all-dielectric systems with extended interaction lengths. Key to our proposal is the use of hollow-core nanofiber (HCNF) and waveguide geometries that simultaneously support a ponderomotive guiding field, enabling long interaction lengths of the order of centimeters (Fig.~\ref{fig:concept}(b)). This method allows us to predict record-high coupling values reaching $|g_Q|=16.07$ (corresponding to over 250 photons emitted per electron) for transmission electron microscopy (TEM) energies ($E=200$~keV). For scanning electron microscopy (SEM) electron energies ($E=17.8$~keV), and using Bragg HCNFs, we show that in addition to strong coupling ($|g_Q|=2.77$), we predict a large nonlinear phase shift of $\delta_{\mathrm{NL}} \approx 16\pi$, even for electron energies well-exceeding $1~\mathrm{keV}$ and for a photon energy of 2.93 eV. This effect stems from an exceptionally-high Kerr-like nonlinearity emergent in our system Hamiltonian, as well as the long interaction length. We show that the proposed system demonstrates favored emission of photons into a single guided mode family. We then analyze the multimode, quantum-optical dynamics in our system, and identify experimental observables that unveil the single-photon blockade effects. Our results pave the way towards new capabilities in quantum optics, from macroscopic nonclassical light generation to room-temperature, electron-mediated single-photon nonlinearities (Fig.~\ref{fig:concept}(c)). 

We recently became aware of several related works \cite{Xie2024MaximalPhotons,DiGiulio2024TowardModes,Zhao2024UpperPhotons} on the free-electron--photon quantum coupling, that have come out in parallel to this work. 

\section{Hollow-core nanofibers and waveguides as a platform for free-electron-light interactions}

Quantum optics in hollow-core fibers and slot waveguides is a well-established field, where the main focus has been on atomic emitters, positioned or trapped inside the hollow core or slot~\cite{Goban2015SuperradianceWaveguide, Epple2014RydbergFibres, Shahmoon2011StronglyWaveguides, Ito1995OpticalFiber}. For free electrons, methods for guiding electron beams inside microwave ponderomotive waveguides \cite{Zimmermann2021ChargedChip, Hammer2015MicrowaveElectrons}, free space \cite{Schachter2020ElectronBeam}, or dielectric laser accelerators \cite{England2014DielectricAccelerators,Hanuka2017TrappingAccelerators,Shiloh2022MiniatureControl,Shiloh2021ElectronAcceleration} have been proposed and implemented. In the latter case, the aim of electron guiding is to mitigate the effect of acceleration-induced diverging forces acting on the electrons (which do not occur in the settings considered in this paper). To the best of our knowledge, electron guiding has never been proposed for the enhancement of quantum light emission and single photon nonlinearity, as detailed in this work.  

We consider the two structures presented in Fig.~\ref{fig:concept}(d), consisting of either a uniform HCNF, with inner radius $a$, outer radius $b$ and refractive index $n$; or a Bragg HCNF, with refractive indices $n_1,n_2$, a periodicity $\Lambda$ and duty-cycle $D$. Uniform HCNFs are more suitable for higher electron energies (above 80~keV, as in TEMs), whereas periodic structures enable us to extend the range of efficient coupling to lower electron energies (down to few-keV, as in SEMs), by slowing down the phase velocity of light \cite{GarciaDeAbajo2010,Roques-Carmes2023Free-electronlightNanophotonics}. These two examples also allow us to investigate two different types of free-electron--photon phase matching, as illustrated in Fig.~\ref{fig:concept}(e): the conventional case of an \textit{intersection} point between the electronic and photonic dispersion curves [defined, respectively, by the RHS and LHS in Eq. (\ref{eq:phase-matching})], and the previously unexplored case of a \textit{tangency} point. As we show below, the scaling of the coupling strength $|g_Q|$ with the interaction length and the nature of the single-photon nonlinearity will significantly differ between these two configurations. 

\begin{figure*}
    \centering
    \includegraphics[scale = 0.75]{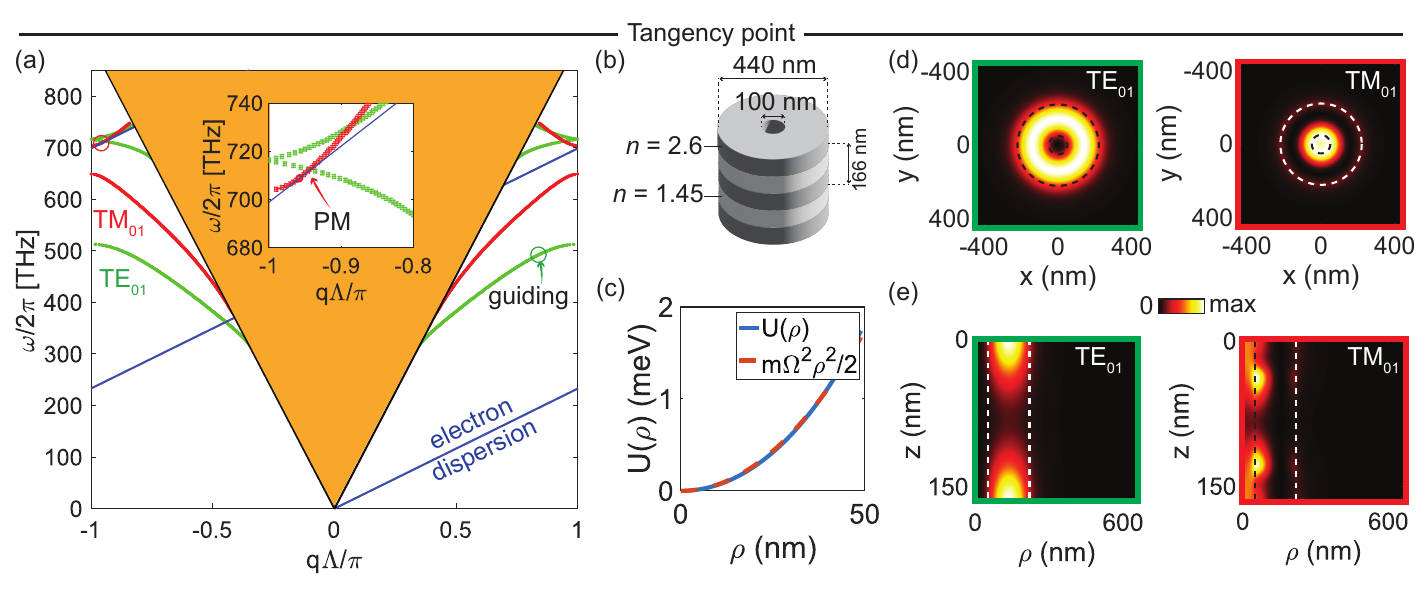}
    \caption{\textbf{Tangency phase-matching in free-electron ponderomotive fiber guiding.} (a) TE and TM band structure of a hollow Bragg nanofiber design. (b) Schematic of the corresponding design. (c) Corresponding ponderomotive potential from the TE mode with peak pump power $P_0 = 500~\text{W}$. (d) Transverse cross sections of mode profiles for TE ($|E_\varphi|$) and TM ($|E_z|$) modes. (e) Longitudinal cross sections of mode profiles for TE ($|E_\varphi|$) and TM ($|E_z|$) modes. In (d) and (e), dashed lines indicate inner and outer fiber boundaries. PM: phase matching. Parameters used in the calculation are $a=50~\mathrm{nm}$, $b=220~\mathrm{nm}$, $\Lambda=165.7~\mathrm{nm}$, $D=32.48\%$, with $n_1=2.6$ (achievable with SiC nanophotonic waveguides \cite{Yi2022SiliconPhotonics} or chalcogenide glass nanofibers \cite{Tezuka2016Mid-infraredFiber}), $n_2 = 1.45$ (using SiO2 or, alternatively, air corrugations), and electron energy of $E=17.8\mathrm{keV}$. Note that, in the periodic case, the electron dispersion line in (a) wraps around the Brillouin zone. }
    \label{fig:bandstructures-tangency}
\end{figure*}

HCNFs support several guided mode families~\cite{Ito1995OpticalFiber}: TM modes, with a purely transerve magnetic field, TE modes, with a purely transverse electric field, and HE and EH hybrid modes, where both electric and magnetic fields have longitudinal components. Note that TM, HE and EH modes all have longitudinal electric field components. TE modes on the other hand, have zero longitudinal electric field components. The modes are typically labeled by indices $l,p$, standing for the orbital angular momentum (OAM) number and their radial number, respectively. Fig.~\ref{fig:bandstructures-intersection}(a,b) present the TM01 and TE01 bands of a uniform hollow nanofiber, together with the electron dispersion line (see simulation parameters in the caption of Fig.~\ref{fig:bandstructures-intersection} and Appendices C and D). Mode profiles corresponding to the highlighted points on each band are presented in Fig.~\ref{fig:bandstructures-intersection}(d,e). The second HCNF design we study in this paper is shown in Fig.~\ref{fig:bandstructures-tangency}. Fig.~\ref{fig:bandstructures-intersection}(a,b) present its band diagram, inside the first Brillouin zone, of a Bragg HCNF. TE01 and TM01 mode profiles, corresponding to the highlighted points in Fig.~\ref{fig:bandstructures-tangency}(a), are presented in Fig.~\ref{fig:bandstructures-tangency}(d,e).

Below, we outline the essential properties of our proposed system, and how they enable the enhancement of free-electron--photon coupling, single photon nonlinearity, and $\beta$-factor. This is achieved through the ponderomotive guiding enabled by the TE01 mode, simultaneously with the strong coupling with the TM01 mode.

\section{Ponderomotive guiding of free electrons in HCNFs}

We consider a free-electron wavepacket propagating inside the hollow core, while being guided by a co-propagating trap potential: an intense laser pulse occupying the TE01 mode supported by the structure (see Figs.~\ref{fig:bandstructures-intersection}(c) and~\ref{fig:bandstructures-tangency}(c) for TE01 mode profiles). The guiding potential is induced through the elastic ponderomotive interaction (the $A^2$ term in the minimal-coupling Hamiltonian) felt by the electron, and, inside the hollow core $\rho<a$ ($a$ is the radius of the hole), takes the form: 
\begin{equation}
    U_p(\rho)=\frac{e^2}{m_e\omega^2_{\mathrm{TE}}}|E_{\mathrm{TE}}(\rho)|^2\approx\frac{1}{2}m_e\Omega^2\rho^2
    \label{eq:ponderomotive-trapping}
\end{equation}
where $e,m_e$ are the electron charge and mass, respectively, and $\omega_{\mathrm{TE}}, E_{\mathrm{TE}}$ are the TE frequency and electric field, respectively. The radial dependency of the electric field of the chosen TE01 mode is given by the first-order Bessel function $J_1(\zeta \rho)$ for $\rho<a$, where $\zeta$ is a complex vacuum transverse wavevector. Thus, for $|\zeta a|\ll 1$, Eq.~(\ref{eq:ponderomotive-trapping}) is well approximated by a 2D parabolic potential, as given by the right hand side, with $\hbar\Omega$ denoting the ground state energy (the parabolic fitting is shown in Figs.~\ref{fig:bandstructures-intersection}(c) and~\ref{fig:bandstructures-tangency}(c)). As a result, a set of guided electron wavefunctions can be supported by the structure, with the fundamental Gaussian mode $\psi_{0,0}(\rho)$ having spatial uncertainty $\Delta r_{e}=\sqrt{\hbar/2m_e\Omega} \ll a$. Higher-order guided wavefunctions, $\psi_{l,p}(\rho)$ can also be supported.

Assuming the electron and trap pulse co-propagate, classical theory predicts that a portion of the electron phase-space trajectories can be guided indefinitely; this is in contrast to the non-guided case where almost all electrons eventually hit the structure walls. Considering a quantum treatment, where electrons can also tunnel outside the trap into the surrounding material, the mean free path of the guided electron mode can be calculated using leaky-mode theory (see Appendix B). The resulting tunneling mean free path can still be orders of magnitude larger than other relevant limiting length scales, such as the propagation loss length, or the group-velocity mismatch (GVM) length, $L_{\mathrm{GVM}} = \tau/|1/v_g - 1/v_e|$, between the electron (velocity $v_e$) and the trap (group velocity $v_g$, pulse length $\tau$). Therefore, in our example structures, we choose to limit the interaction lengths according to the latter quantities. The chosen parameters for the TE ponderomotive trap, for each of the examples in this work, are outlined in Appendix~C.

\section{Strong free-electron--photon coupling with large $\beta$-factors}

The free electron guided by the ponderomotive potential, as discussed in the previous section, can be used to achieve strong free-electron--photon coupling. For this purpose, the photon must be in a mode that supports a longitudinal electric field component, and therefore, must be in either the TM, HE or EH mode families. Achieving strong free-electron--photon coupling typically requires satisfying the phase-matching condition
\begin{equation}
    \omega(q)=\frac{E(k)-E(k-q-2\pi m/\Lambda)}{\hbar},
    \label{eq:phase-matching}
\end{equation}
which is a resonance condition between the electron and photon enabling efficient exchange of energy. In Eq.~(\ref{eq:phase-matching}), $E(k)$ is the electron energy at momentum $\hbar k$, $\omega(q)$ is the photonic dispersion relation, $q$ is the photonic wavevector in the waveguide, and $m$ is an integer. For a uniform system, $m=0$, while for a periodic system, $m$ corresponds to the number of times the electron dispersion wraps around the Brillouin zone (Fig.~\ref{fig:bandstructures-tangency}(a)). Graphically, a solution $(q,\omega(q))$ to Eq.~(\ref{eq:phase-matching}) corresponds to a phase-matching point where the electron dispersion curve crosses a photonic band. The two types of phase-matching points, illustrated in Figs.~\ref{fig:bandstructures-intersection} and~\ref{fig:bandstructures-tangency}, are obtained for the two types of structures: for the uniform HCNF, we have an intersection point with the TM01 mode (highlighted in Fig.~\ref{fig:bandstructures-intersection}(a)), for zeroth-order ($m=0$) phase-matching. For the Bragg structure, we obtain a tangency point at the bottom of the TM01 upper band (highlighted in Fig.~\ref{fig:bandstructures-intersection}(f) and its inset), using second-order ($m=2$) phase-matching.  

Strong free-electron--photon coupling manifests as inelastic scattering (i.e., exchange of many quanta of energy and momentum) between the electron and incident quantum light. If no light is initially present, strong coupling corresponds to the spontaneous emission of photons by a single electron \cite{Kfir2019EntanglementsRegime, DiGiulio2019ProbingElectrons, Adiv2023ObservationRadiation}. The latter scenario will be the focus of this work, as it can be used for shaping and heralding of quantum states of light \cite{BenHayun2021ShapingElectrons, Dahan2023CreationElectrons, Feist2022Cavity-mediatedPairs, Huang2023Electron-PhotonCircuits}. In the linear regime ($\delta_{\mathrm{NL}}\ll2\pi$), each electron emits on average $|g_Q|^2$ photons, in a Poissonian distribution, while in the nonlinear regime $\delta_{\mathrm{NL}}\gg2\pi$, we should expect a sub-Poissonian distribution, and vacuum Rabi oscillations as a function of $|g_Q|$ \cite{Karnieli2023Jaynes-CummingsPhotons}. 

We calculate the total coupling efficiency to a general mode family (labeled by $s=\mathrm{TM}_{ij},\mathrm{TE}_{ij},\mathrm{HE}_{ij}$), for $m$-th order phase-matching and for the two types of phase-matching points, as
\begin{equation}
\begin{split}
    |g_{Q,s}|^2 &=\frac{\alpha}{\tilde{A}} \frac{L_{\mathrm{int}}}{\lambda}\Big|\int d^2\boldsymbol{\rho}~\psi_f^*(\boldsymbol{\rho})\psi_i(\boldsymbol{\rho})u_{m,z}^{(s)}(\boldsymbol{\rho})\Big|^2 \\& \times  \begin{cases}
    1/|1-v_{g,s}/v_e| &\text{intersection}\\
    (4/3\sqrt{\pi})\sqrt{L_{\mathrm{int}}/v_{g,s}^2|\beta_2|} &\text{tangency},
    \end{cases}       
\end{split}
\label{eq:gQu_maintext}
\end{equation}
where $\alpha\approx1/137$ is the fine structure constant, $\tilde{A} = A/\lambda^2$ is the normalized mode area, and $\lambda$ is the free-space wavelength of the mode. Here, $\mathbf{u}_m(\boldsymbol{\rho})$ is the $m$-th Fourier component in the expansion of the mode function (a Bloch function in a periodic system), written as $\mathbf{u(\mathbf{r})}=\sum_m \mathbf{u}_m(\boldsymbol{\rho}) e^{i2\pi m z/\Lambda}$. The normalization of $\mathbf{u}(\mathbf{r})$ is such that $\mathbf{E}(\mathbf{r})=\sqrt{\hbar \omega(q)/2\epsilon_0 A} e^{iqz} \mathbf{u}(\mathbf{r})$. $\psi_{i/f}$ denote the initial and final free-electron transverse wavefunctions (here, unless stated otherwise, $\psi_i=\psi_{00}$ is the fundamental mode). Finally, $v_e$ is the electron velocity, and $\beta_2$ is the group velocity dispersion at the phase-matching point. More details on the derivation are provided in Appendix~D.

\begin{figure} [h!]
    \centering
    \includegraphics[scale = 0.5]{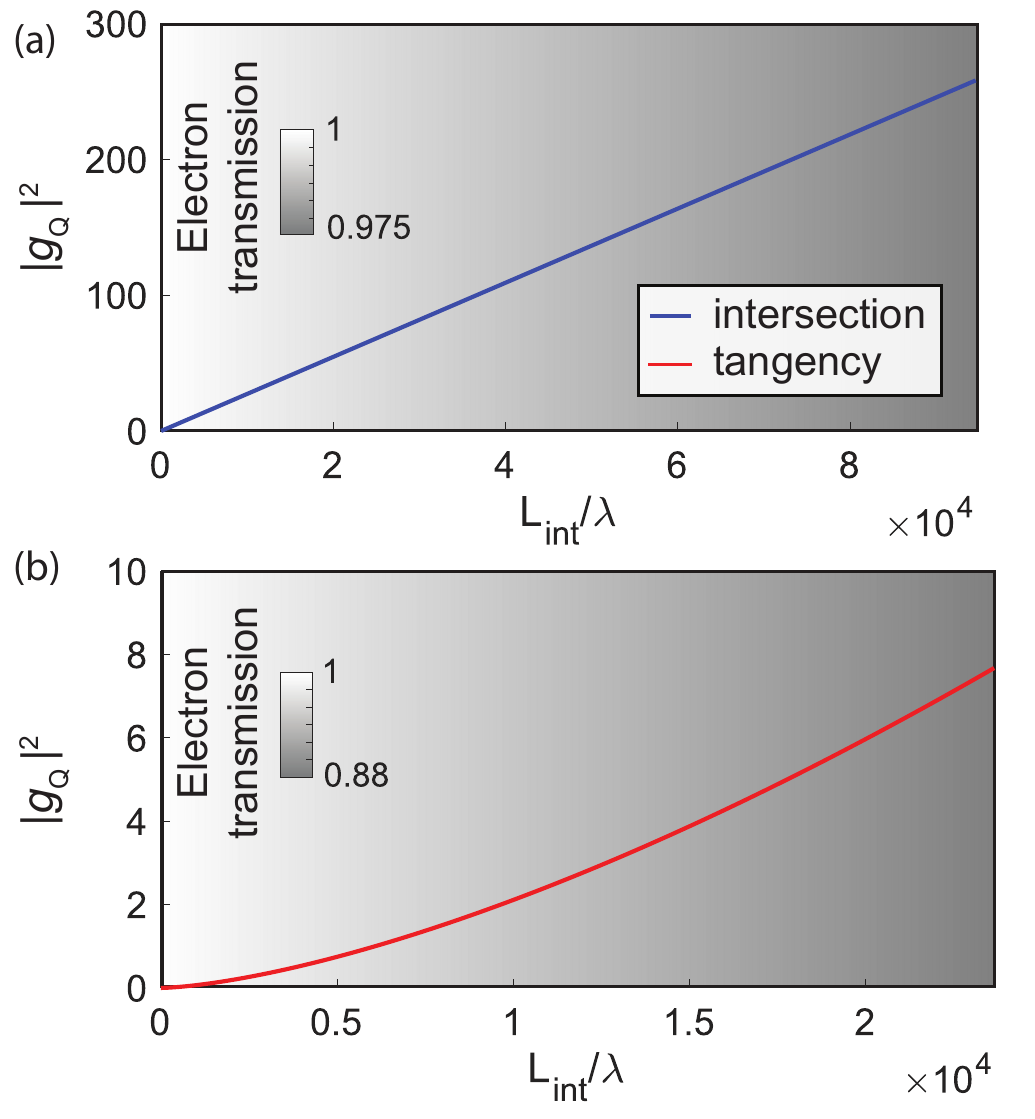}
    \caption{\textbf{Electron-photon coupling: scaling and physical limits.} Scaling of the free-electron--photon coupling as a function of the interaction length is shown for intersection phase-matching (top, blue) and tangency phase-matching (bottom, red). Background color corresponds to the electron transmission probability calculated according to Appendix~B with $-2\text{Im}\left( 2p+1 \right)=10^{-3}$.}
    \label{fig:gqu-perspective}
\end{figure}

From the form of Eq.~(\ref{eq:gQu_maintext}), it is clear that modes with high transverse confinement $\tilde{A}\ll1$ and large overlap with the electron wavefunctions are favored. The type of phase-matching point quantifies the effective number of longitudinal (spectral) modes within the phase-matching bandwidth. Interestingly, this further implies different scalings of $|g_Q|^2$ with respect to the interaction length: $|g_Q|^2\propto L_{\mathrm{int}}$ for an intersection point, and $|g_Q|^2\propto L_{\mathrm{int}}^{3/2}$ for a tangency point. Fig.~\ref{fig:gqu-perspective} shows the different scaling and achievable values of coupling with reasonable experimental parameters. The combination of strong transverse confinement and large overlap of the TM01 mode, together with the guiding-enhanced interaction lengths, enables exceptionally high predicted coupling values of $|g_{Q,\mathrm{TM}}| = 16.07$ and $|g_{Q,\mathrm{TM}}| = 2.77$, for the uniform and Bragg examples, respectively.

Another important figure of merit is the $\beta$-factor, extensively used in waveguide quantum electrodynamics \cite{Sheremet2023WaveguideCorrelations}, defined as the ratio between the coupling to the desired mode (here, TM01), to the total coupling:
\begin{equation}
    \beta = \frac{|g_{Q,\mathrm{TM01}}|^2}{|g_{Q,\mathrm{TM01}}|^2 + \sum_{s\neq \mathrm{TM01}}|g_{Q,s}|^2},
\end{equation}
where the modes $s\neq \mathrm{TM01}$ are all other modes the electron may couple to. Since the TM01 mode is the only one to have a non-zero amplitude at the center of the hollow core, it most efficiently couples to the free electron. Moreover, when the electron wavefunction is spatially confined ($\Delta r_e \ll a$), an electron transition where $\psi_f = \psi_i = \psi_{00}$ is strongly favored. The electron also excites HE and EH modes, as these carry an $E_z$ component, which is along the propagation direction of the electron. However, such modes possess orbital angular momentum (OAM), with a singularity at the center of the hollow core. This will reduce the overlap with the transverse electron wavefunction, and further requires that the electron exchanges an OAM quantum with the field. Thus, $\psi_f$ must have a nonzero OAM quantum number, whereas its largest overlap with $\psi_i=\psi_{00}$ happens for a radial number $p=0$. Eventually, this overlap can be made arbitrarily small using stronger electron transverse confinement. Thus, we can ensure that the $\beta$-factor approaches unity, even when the coupling efficiency is arbitrarily increased~\footnote{We note that the $\beta$-factor for free-electron--photon coupling was also named ``coupling ideality'' in the literature \cite{Huang2023Electron-PhotonCircuits}.}. In our examples, we find $\beta$-factors of $0.89$ (resp., $0.99$) for the uniform (resp., Bragg) designs. For the Bragg design, coupling to $l\neq 0$ modes is further suppressed by the strongly-decaying evanescent fields that accompany the nonzero Fourier components of the Bloch functions.

\begin{figure*}
    \centering
    \includegraphics[scale = 0.45]{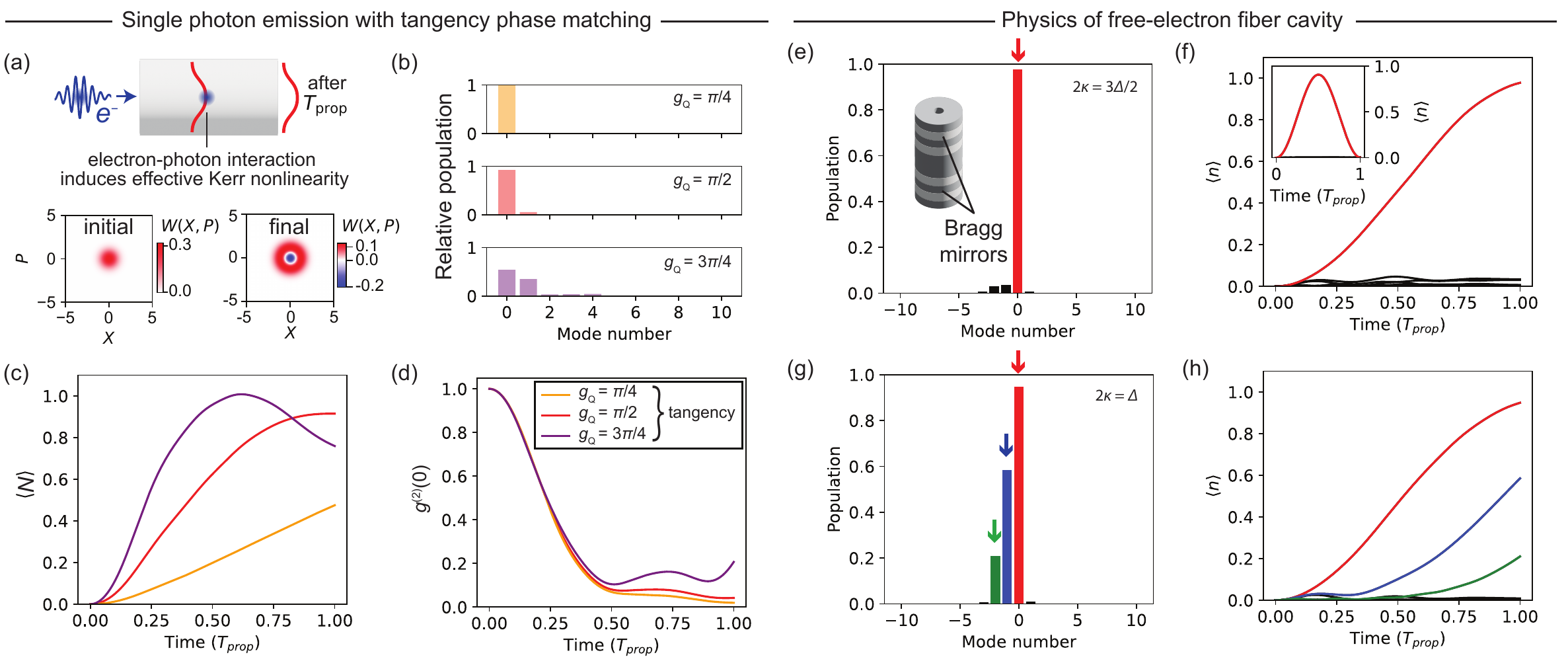}
    \caption{\textbf{Single-photon nonlinearity and cascaded photon emission in free-electron photonic fibers and fiber cavities.} (a-d) Tangency phase matching. (a) Deterministic single photon emission in free-electron--photonic fiber interactions. Insets: initial state Wigner function (left); simulated final state Wigner function (right), for $|g_Q|=\pi/2$. (b) Simulated occupations on supermode basis, for different values of $|g_Q|$. (c) Time evolution of the total photon number for different values of $|g_Q|$ as in (b). (d) Time evolution of the second-order correlation function at zero-delay. (e-f) Single-photon nonlinearity in free-electron fiber cavities. (e) cavity mode population for $|g_Q|=\pi/2$ and $2\kappa = 1.5\Delta$ and $E=3.72$ keV. (f) Time evolution of cavity mode photon numbers (with same parameters as in (e)), recovering the Jaynes-Cummings regime for $E=3.72$ keV. Inset: For reference, the time evolution showing an almost ideal, full Rabi cycle for $|g_Q|=\pi$ and $E = 100~$eV is plotted. (g,h) Same as in (e,f) with $E=3.72~\mathrm{keV}$, but for $2\kappa =\Delta$, resulting in cascaded emission into consecutive cavity modes.}
    \label{fig:tangency}
\end{figure*}


\section{Kerr-like Hamiltonian and single-photon nonlinearity enabled by free electrons}
In the previous sections we have seen how extending the interaction length via electron guiding can dramatically increase the coupling. Intriguingly, an extended interaction enables us to enter a completely different physical regime where quantum recoil corrections~\cite{Huang2023QuantumLattices} can manifest as a \textit{single photon nonlinearity} mediated by free electrons. 

The physical intuition for this phenomenon is as follows: for swift electrons and short interaction lengths, the electron experiences no recoil (or back-action) by an emitted photon. Thus, an electron already phase-matched to an optical mode remains phase-matched even after emitting a photon, and can, therefore, continue emitting more photons. The interaction length is assumed to be short enough such that even if such recoils do occur, they do not accumulate to any significant effect. 

However, the situation is very different when we consider slower electrons and long interaction lengths: slower electrons can experience a non-negligible recoil after the emission of even a single photon, such that after the emission their momentum has changed. This results in an electron detuning from the optical mode (e.g, if the electron was originally phase matched, after emitting one photon, it is no longer phase matched): and in that case, the longer the interaction length, the larger the effect this detuning has on a consecutive emission of a second photon. In the limit where the detuning-induced phase mismatch accumulated along the interaction becomes very large, the electron cannot emit a second photon at all.    

To elucidate this single-photon nonlinearity mediated by free electrons, we first write the non-relativistic free-electron--photon interaction Hamiltonian as
\begin{equation}
    H = \frac{p^{2}}{2m} + \sum_{q}^{}{\hslash\omega_{q}a_{q}^{\dagger}a_{q} + \sum_{q}^{}{\hslash\Omega_{q}(a_{q}b^{\dagger}_q + a_{q}^{\dagger}b_q),}}
\end{equation}
where $p$ is the electron momentum, \(\Omega_{q} = ev_{0}\sqrt{\hslash/2\epsilon_{0}\omega_{q}V}\) is the coupling constant for each mode (satisfying $|g_Q|^2=\sum_q |\Omega_q|^2 \mathrm{sinc}(\Delta_q L_{\mathrm{int}}/2v)$, with the detuning $\Delta_q$ defined below), and $b_q = e^{-iqz}$ is the electron momentum ladder operator, with $[b_q,b_{q'}^{\dagger}]=0$ \cite{Kfir2019EntanglementsRegime}. In Appendix F, we show that in a subspace with a fixed total momentum $P$, this Hamiltonian can be recast into the form of an effective Kerr nonlinearity acting on an effective photon:
\begin{align}
    H_{\mathrm{eff}} = \sum_{q}\hbar \Delta_q A_{q}^{\dagger}A_{q}  + \sum_{q}^{}{\hslash\Omega_{q}(A_{q} + A_{q}^{\dagger})} + \hbar\kappa N(N-1),    \label{eq:kerrhamiltonian-maintext}
\end{align}
where $\kappa = \hbar q_0^2 / 2m$ is the effective Kerr nonlinearity around the phase-matching point $(q_0,\omega_0)$, and  
\begin{align}
\Delta_q=\omega_{q} - qv +\frac{\hbar q^2}{2m}
\end{align}
denotes the free-electron--photon detuning. The total excitation number is given by:
\begin{align}
N = \sum_{q}^{}{A_{q}^{\dagger}A_{q}}.
\end{align}
In writing Eq.~(\ref{eq:kerrhamiltonian-maintext}), we have introduced the hybrid ladder operators $A_{q}  = a_{q}b_{q}^{\dagger}$, which act on a joint Hilbert subspace of the free electron and photons, spanned by the states $\ket{P-n\hbar q}_{\mathrm{el}}\ket{n_q}_{\mathrm{ph}}$, where $A_q\ket{P-n_q\hbar q}_{\mathrm{el}}\ket{n_q}_{\mathrm{ph}}=\sqrt{n_q}\ket{P-(n_q-1)\hbar q}_{\mathrm{el}}\ket{n_q-1}_{\mathrm{ph}}$. Using the commutation relations of \(b_q\), we find that {[}\(A_{q},A_{q'}^{\dagger}\rbrack = \delta_{q,q'}\): these new operators have the same commutation relations as photon operators, and further satisfy $A^{\dagger}_q A_q = a^{\dagger}_q a_q$ (the latter holds for a single-electron system). With that in mind, we see that Hamiltonian Eq.~(\ref{eq:kerrhamiltonian-maintext}) effectively describes a multimode bosonic system with a Kerr nonlinearity that couples to the total excitation number. For spontaneous emission, where the initial electron momentum is $p=mv$ and the electromagnetic field is initially at vacuum, a phase-matched electron could excite a (generally multimode) state. The Kerr term $\hbar \kappa N(N-1)$ ensures that any excitation number beyond $N=1$ is detuned from phase matching, and in the limit where this detuning is large enough, we should expect a photon blockade where further light emission by the electron is suppressed. 

The inherent multimode nature of the problem (namely, the summation over $q$ in Eq. (6)) complicates the qualitative description we gave in the beginning of the section. In fact, in a continuous multimode system, such as the guided-wave structures considered here, the type of phase-matching point can significantly affect the dynamics. For example, for an intersection point phase-matching in a continuous case, a nonlinear detuning as in the Hamiltonian from Eq.~(\ref{eq:kerrhamiltonian-maintext}) can indeed cause a considerable phase mismatch between the free electron and an initially emitted photon wavepacket. Graphically, the detuning is manifested as a change in the slope of the electron dispersion - since recoil changes the electron velocity - and as such, the electron dispersion no longer intersects the photon dispersion at the same point. However, as illustrated in Fig.~\ref{fig:concept}(e) (left panel), there exists a continuum of other available intersection points. Thus, although high nonlinearity suggests that an electron tends to not emit twice into the same mode, it can in fact continue emitting more photons into other modes, in a cascaded fashion. While this observation implies that a single-photon blockade effect is less likely to happen at a continuous intersection phase matching point, it can still be of practical use. For example, by partially filtering some of the modes, it may still be possible to obtain a photon-number state without post-selection of the electron energy.  

The situation is very different if one considers instead a tangency phase-matching point: in this case, a single recoil leaves the electron dispersion curve below the photonic dispersion curve (Fig.~\ref{fig:concept}(e), right panel), and thus the electron stays detuned from any other modes after a single excitation. We therefore focus our attention on this type of dynamics. We note, however, that the multimode nature of the dynamics still requires a more complicated treatment than in the linear regime, as the electron can interact with multiple \textit{supermodes}: photonic wavepacket modes with different spectral (or temporal) shapes. In Appendix~G, we develop a nonlinear supermode formalism that captures the dynamics in our system, and use it to calculate relevant experimental observables. We find that for values of $g_Q$ as high as $\pi/2$ (a half vacuum-Rabi oscillation), the free electron mainly excites a single fundamental supermode, which allows us to further evaluate the resulting quantum state contained in that supermode.   

Figs.~\ref{fig:tangency}(a-d) depict simulation results for the tangency phase-matching point in the nonlinear regime, corresponding to the parameters of the tangency point example in Fig.~\ref{fig:bandstructures-tangency} (see Appendix~D for more details), and for different values of the total coupling $|g_Q|$. The predicted single-photon Kerr nonlinearity for this setup is $\kappa= 2\pi \times 30.06~\text{GHz}$, corresponding to a nonlinear phase $\delta_{\mathrm{NL}} =2\kappa L_{\mathrm{int}}/v \approx 16\pi$. Fig.~\ref{fig:tangency}(a) shows the calculated Wigner functions occupying the fundamental supermode, for $|g_Q|=\pi/2$. As expected, the Wigner function shows a considerable negativity, and resembles that of a single-photon Fock state $W(\alpha)= (2/\pi)\exp(-2|\alpha|^2)(4|\alpha|^2-1)$, a donut-shaped distribution with a negative central dip. Fig.~\ref{fig:tangency}(b) shows the corresponding supermode populations for different values of $|g_Q|$, and Fig.~\ref{fig:tangency}(c) shows the time evolution of the expectation value of the total excitation number, for the same values of $|g_Q|$ as in Fig.~\ref{fig:tangency}(a,c). The total excitation number experiences a vacuum Rabi oscillation, though with limited visibility, owing to the excitation of other supermodes (interestingly, a similar behavior is obtained in the deep-quantum regime of multimode parametric downconversion \cite{Jankowski2024UltrafastDynamics}). Finally, Fig.~\ref{fig:tangency}(d) shows the value of the second-order correlation function at zero delay, $g^{(2)}(0)=\braket{N(N-1)}/\braket{N}^2$, as a function of interaction time and for different total coupling $|g_Q|$ (note that here we evaluate $g^{(2)}(0)$ with respect to the \textit{total} excitation number $N$). A value of $g^{(2)}(0)\ll 1$ at the end of the interaction indicates that the system approximately remains in the single-excitation manifold, which is a hallmark of the expected single-photon nonlinearity. For $|g_Q|>\pi/2$, coupling to supermodes other than the fundamental decreases this effect. We emphasize that these strong nonlinear effects are obtained even for energies well-exceeding $1~\mathrm{keV}$ (namely, $17.8~\mathrm{keV}$ in this example), which are more experimentally accessible than the sub-keV energies that have been recently considered for quantum recoil effects \cite{Talebi2020StrongPrinciples,GarciaDeAbajo2022CompleteElectrons,Eldar2024Self-TrappingDomain,Karnieli2023Jaynes-CummingsPhotons, Pan2023Low-energyApplications, Synanidis2024QuantumLight}. 

Interestingly, other regimes of interaction emerge if the photonic system hosts a discrete set of modes, e.g., as in a fiber cavity, which could be realized in our system by introducing Bragg mirrors at fiber ends (Fig.~\ref{fig:tangency}(e), inset). In this case, the electron dispersion can intersect the photonic modes only at certain, equally-spaced values. Now, when the electron experiences recoil following emission of a photon into cavity mode $j$, it may or may not be detuned from other neighbouring modes: this will strongly depend on the ratio between the recoil detuning $2\kappa$ and the effective spacing between the modes (defined in Eq.~\ref{effFSR} in Appendix~G), which we call the effective free spectral range and denote by $\Delta$. 

A specific choice of group velocity mismatch between the electron and cavity modes makes the cavity supermode basis coincide with the original set of cavity modes (more details can be found in Appendix~G). In this manner, the fundamental supermode excited by the electron becomes a \textit{single} cavity mode. Two distinct behaviors are depicted in Fig.~\ref{fig:tangency}(e-h). For a choice of nonlinearity $2\kappa = 1.5 \Delta$, the electron excites a single cavity mode and is detuned from the rest, as can be seen in the cavity mode population (Fig.~\ref{fig:tangency}(e)) and the evolution of the average photon numbers (Fig.~\ref{fig:tangency}(f)), displaying a distinct vacuum Rabi oscillation. This is the so-called Jaynes-Cummings regime~\cite{Karnieli2023Jaynes-CummingsPhotons}, where the electron reduces to an effective two-level emitter interacting with a single mode. Another choice, this time of $2\kappa = \Delta$, makes sure that following every recoil, the electron is always phase-matched to a neighboring mode. This results in a cascaded emission of photons into several cavity modes, as can be seen in the mode population (Fig.~\ref{fig:tangency}(g)) and photon number dynamics (Fig.~\ref{fig:tangency}(h)). We note again that these effects can occur at energies higher than $1~\mathrm{keV}$ ($3.72~\mathrm{keV}$ in this example).

\section{Discussion and outlook}

We have shown that ponderomotive guiding in hollow-core photonic structures such as fibers and waveguides, could be used to increase the interaction length between free electrons and photons. As a result, we predict that such structures could strongly enhance both free-electron--photon coupling and single-photon nonlinearities mediated by free electrons.

While several platforms are promising candidates to achieve strong free-electron--photon coupling~\cite{Adiv2023ObservationRadiation, Feist2022Cavity-mediatedPairs,Bezard2023HighCavities, DMello2023EfficientWaveguide, Huang2023Electron-PhotonCircuits, Yang2023PhotonicRadiation}, these proposals are ultimately hampered by electron diffraction, limiting the interaction length between free electrons and photons. Ponderomotive guiding, on the other hand, can result in orders-of-magnitude increase in free-electron--photon coupling in either intersection or tangency phase matching designs. Tangency phase matching has the additional advantage of matching the group velocity of the free-electron wavepacket to that of the emitted photonic mode and that of the guiding field. In principle, one could also achieve group velocity matching between the electron and the guiding field, even in intersection phase matching, with appropriate dispersion engineering. This is favorable for increasing the interaction length and decreasing the required guiding optical fluence. Further, in our current design, the guiding field is mostly confined in the dielectric structure, which results in limited guiding power due to damage threshold of the material. Alternative designs employing dispersion engineering and allowing higher optical powers could rely, e.g., on air guiding in photonic crystal fibers~\cite{Temelkuran2002Wavelength-scalableTransmission, Joannopoulos2008PhotonicEdition.}, thereby alleviating these limitations.

The long interaction lengths predicted in our designs would enable the realization of strong single-photon nonlinearities using free electrons. The designs proposed in this paper also rely on purely dielectric materials, thereby mitigating the effect of losses on quantum dynamics and coherence. The realization of single-photon nonlinearities with free electrons opens up many exciting prospects in free-electron quantum optics, such as the efficient and deterministic (or heralded) generation of non-Gaussian light. In the proposed system, owing to the availability of a large coupling coefficient $g_Q$, such effects can in principle be observed even after a single pass of a single free electron, instead of slower, multi-electron consecutive interactions previously considered in the literature~\cite{Dahan2023CreationElectrons, BenHayun2021ShapingElectrons}. As we showed in Fig.~\ref{fig:tangency}, the strong nonlinearity induced by free electrons also results in complex multimode dynamics that are reminiscent of broadband parametric down conversion in the deep quantum regime~\cite{Jankowski2024UltrafastDynamics}.

While we expect that the realization of strong coupling between free electrons and photons to be an immediate application of our ponderomotive guiding design, it is worth noting that atomic emitters could also be integrated in the hollow core of the fiber~\cite{Goban2015SuperradianceWaveguide, Epple2014RydbergFibres, Shahmoon2011StronglyWaveguides, Ito1995OpticalFiber}. With recent proposals highlighting the possibility of controlling atomic emitters with modulated electron beams~\cite{Gover2020Free-Electron-Bound-ElectronInteraction}, we expect that this platform could also be utilized to explore fascinating and complex quantum dynamics involving three flavors of interacting quantum systems: free electrons, photons, and atomic emitters \cite{Karnieli2023QuantumElectrons,Lim2023QuantumWavefunctions,ArqueLopez2022AtomicElectrons}.

\setcounter{figure}{0}
\setcounter{equation}{0}

\renewcommand{\thefigure}{S\arabic{figure}}
\renewcommand{\theequation}{S\arabic{equation}}

\section{Appendix A : Geometrical bound on free-electron--photon coupling strength}
In this Appendix, we outline a simple geometric argument showing the limitation on the interaction length due to electron beam diffraction. From Fig.~\ref{fig:concept}(a), we can estimate the interaction length as $L_{\mathrm{int}}=2h/\theta$. Since $|g_Q|^2\propto L_{\mathrm{int}} \exp{(-2h/\lambda_{\mathrm{ev}})} \propto h \exp{(-2h/\lambda_{\mathrm{ev}})}$, where $\lambda_{\mathrm{ev}}$ is the evanescent wavelength, the optimal positioning of the e-beam above the sample is $h_{\mathrm{opt}}=\lambda_{\mathrm{e}}/2$. From phase-matching considerations, $q_z = \omega/v_e = 2\pi / \lambda\beta_e$ (with $\beta_e=v_e/c$), and $\sqrt{q_z^2-q^2}= 1/\lambda_{\mathrm{ev}}$, we can write $\lambda_{\mathrm{ev}} = \beta_e \gamma_e \lambda/2\pi$ (with $\gamma_e = 1/\sqrt{1-\beta_e^2}$), and find that the interaction length is bounded by
\begin{equation}
    \begin{split}
        L_{\mathrm{int}} &= \frac{2h}{\theta}\leq 2\pi\beta_e\gamma_e \frac{hw_0}{\lambda_{\mathrm{C}}}\\
        &= \frac{\beta^3_e\gamma_e^3}{8\pi}\frac{\lambda^2}{\lambda_{\mathrm{C}}} \left(\frac{h}{h_{\mathrm{opt}}}\right) \left(\frac{w_0}{h_{\mathrm{opt}}}\right),
    \end{split}
\end{equation}
where the inequality holds for a Gaussian electron beam of waist $w_0$, $\lambda_{\mathrm{C}}=2.426~\mathrm{pm}$ is the Compton wavelength, and we consider both $w_0$ and $h$ relative to the optimal positioning $h_{\mathrm{opt}}$. With diffraction, there is no obvious way to increase $g_Q$ by increasing $L_{\mathrm{int}}$: if we increase $h$ to increase $L_{\mathrm{int}}$, then $|g_Q|$ decays exponentially. Further, $L_{\mathrm{int}}$ quickly decreases with decreasing electron velocity $\beta_e$. Considering optimal positioning $h=h_{\mathrm{opt}}$, a modest confinement of $w_0=h_{\mathrm{opt}}/2$ and $\lambda = 500~\mathrm{nm}$, we have that $L_{\mathrm{int}}= 2.1~\mathrm{mm}\times \beta^3_e \gamma_e^3$. For the energies considered here, $E=200~\mathrm{keV}$ ($\beta_e=0.7$) and $E=17.8~\mathrm{keV}$ ($\beta_e=0.257$), we find that $L_{\mathrm{int}}\leq 2~\mathrm{mm}$ and $L_{\mathrm{int}}\leq 40~\mathrm{\mu m}$, respectively.

\section{Appendix B: Free-electron mean free path in ponderomotive potential}
In this Appendix, we evaluate the mean free path for electron guiding with a finite parabolic potential. A classical electron trajectory analysis predicts that all electron trajectories with transverse kinetic energy less than or equal to the maximum value of the trap potential are guided indefinitely. However, in a quantum picture, free electrons may tunnel outside of the trap and into the surrounding bulk material. To evaluate the typical tunneling propagation length, we envision the equivalent optical problem of a leaky waveguide, and employ leaky mode theory \cite{Hu2009UnderstandingRevisited}. For this we write the Schrödinger equation in cylindrical coordinates:
\begin{equation}
    -\frac{\hbar^2}{2m}\left[\frac{\partial^2}{\partial z^2} + \frac{1}{\rho} \frac{\partial}{\partial\rho} \left(\rho\frac{\partial}{\partial\rho}\right)\right]\psi+V(\rho)\psi = E\psi,
    \label{eq:schrodinger-trapping}
\end{equation}
with
\begin{equation}
    V(\rho)=\begin{cases} \frac{1}{2}m\Omega^2\rho^2 &\rho\leq a\\
    -eU &\rho>a
    \end{cases}       
\end{equation}
where $U>0$ is the mean inner potential of the bulk material surrounding the hollow core. 

\begin{figure}
    \centering
    \includegraphics[scale=0.25]{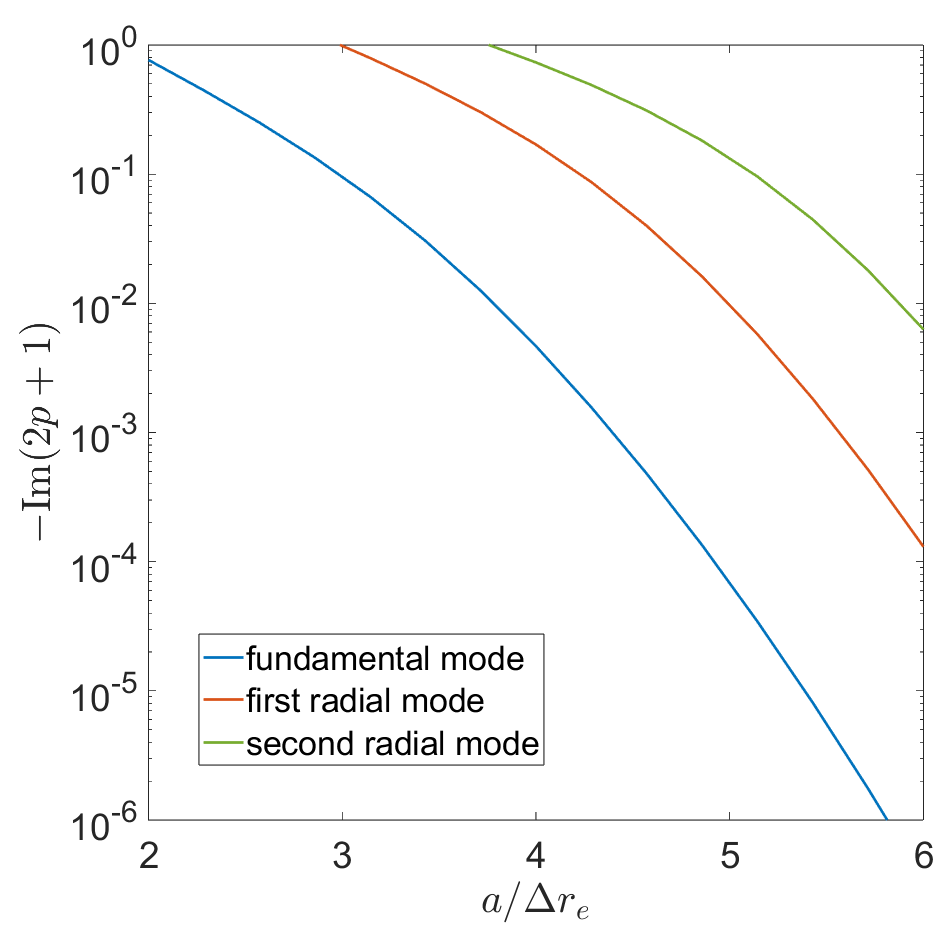}
    \caption{\textbf{Ponderomotive guiding scaling.} Scaling of the inverse electron mean free path (given by Eq.~(\ref{eq:scaling-mfp})) as a function of electron confinement factor for various guided electron modes.}
    \label{fig:guiding}
\end{figure}

In the absence of guiding ($\Omega\to 0$), we consider a worst-case scenario estimate, where an electron impinging on the wall at $\rho=a$ is fully transmitted and thereby lost. Thus, we set $U=0$. In fact, choosing $U>0$ will increase the mean free path, as this results in a nonzero reflection probability back into the hollow core, and can also be shown by numerically solving Eq.~(\ref{eq:schrodinger-trapping}). For simplicity, we solve the Schrödinger equation for radial modes without OAM, so we write $\psi(\rho,z)=e^{ik_z z}\phi(\rho)$, we find that $\phi$ satisfies:
\begin{equation}
    \frac{1}{x}\frac{d\phi}{dx} +\frac{d^2\phi}{dx^2}+\left(2p+1 -\frac{x^2}{4}\Theta(\bar{a}-x) \right) \phi =0       
\end{equation}
where $x=\rho/\Delta r_e$ (with $\Delta r_e=\sqrt{\hbar/2m\Omega}$), $\bar{a}=a/\Delta r_e$, $2p+1 = (E-\hbar^2k_z^2/2m)/\hbar\Omega$, and $\Theta(\bar{a}-x)$ is the Heaviside step function. The only solution that is non-divergent at $x=0$ is of the form
\begin{equation}
    \phi(x) = \begin{cases} A e^{-x^2/4} L_{p}^{0}(x^2/2)&x\leq\bar{a}\\
    BH_0^{(1)}(\sqrt{2p+1}x) &x>\bar{a}, \end{cases}       
\end{equation}
where $L_n^m(x)$ is the generalized Laguerre polynomial, and $H_0^{(1)}(x)$ is the Hankel function of the first kind. In the limit $\bar{a}\to\infty$, $p$ is an integer, denoting the radial index of the guided mode. Otherwise, $p$ is approximately an integer, and may have a small and negative imaginary part. Note that for non-integer orders, the Laguerre polynomials are also given by $L_n^0(x) = {}_1F_1(-n,1,x)$. Requiring that $\phi$ and its derivative are continuous at $x=\bar{a}$, using $(L_n^0(x))'=-L_{n-1}^1(x)$ and $(H_0^{(1)}(x))'=-H_{1}^{(1)}(x)$, we find the dispersion relation for $p$:
\begin{equation}
    \frac{1}{2}+\frac{L_{p-1}^1(\bar{a}^2/2)}{L_{p}^0(\bar{a}^2/2)} = \frac{\sqrt{2p+1}H_1^{(1)}(\sqrt{2p+1}\bar{a})}{\bar{a}H_0^{(1)}(\sqrt{2p+1}\bar{a})}.
    \label{eq:dispersion-ponderomotive}
\end{equation}
We numerically solve Eq.~(\ref{eq:dispersion-ponderomotive}) by searching for its complex roots near $2m+1,~m=0,1,2,...$. Once found, we can evaluate the lifetime via:
\begin{equation}
    k_z=\sqrt{\frac{2m}{\hbar^2}\left[E-\hbar\Omega(2p+1)\right]}\approx k_{z,0}-i\frac{\Omega}{v} \mathrm{Im}(2p+1),
\end{equation}
such that
\begin{equation}
    \Lambda_{\mathrm{MFP}}^{-1}=2\mathrm{Im}k_z \approx - \frac{2\Omega}{v}\mathrm{Im}(2p+1).
    \label{eq:scaling-mfp}
\end{equation}
A similar expression could be obtained using the WKB approximation, where $-\mathrm{Im}(2p+1)$ is interpreted as a tunneling probability, and $2\Omega/v$ is the frequency (in units of propagation length) at which the electron impinges on the barriers (edges of the trap). Fig.~\ref{fig:guiding} shows the values of $-\mathrm{Im} (2p+1)$ as a function of the confinement $\bar{a}=a/\Delta r_e$, for the first three radial modes. Typical values of $v/2\Omega=2\pi(v/c)  (a^2/\lambda_{\mathrm{C}} )(\Delta r_e/a)^2$ are between tens of microns to millimeters, with $a/\Delta r_e \approx ~ 4.5-8$ resulting in extremely long mean free path. However, the latter decreases by orders of magnitude with decreasing confinement, $a/\Delta r_e$.   

Finally, Fig. \ref{fig:guiding} also provides insight into quantifying losses due to coupling mismatch between the electron wavefunction inserted into the fiber and the guided modes. Such mismatch might occur, e.g., due to mismatch in electron beam waist size or due to limited transverse coherence. We can decompose an input wavefunction in the basis of guided modes and estimate such mismatch losses by considering the corresponding reduction in the MFP of the higher-order radial modes.

\section{Appendix C: ponderomotive guiding and maximal interaction length parameters}
In this Appendix, we list the parameters chosen for the ponderomotive traps in each of the examples in the main text. We consider a TE pulse of length $\tau = 100~\mathrm{psec}$ ($0.5~\mathrm{psec}$), peak power $P_0 = 30~\mathrm{W}$ ($500~\mathrm{W}$), at wavelength $\lambda_{\mathrm{TE}} = 1200~\mathrm{nm}$ ($610~\mathrm{nm}$) for the uniform (Bragg) fiber. The position of the modes on the TE01 bands and the mode profiles are illustrated in Fig.~\ref{fig:bandstructures-intersection} and \ref{fig:bandstructures-tangency}. 

The maximal fluences of the TE mode we use are $0.9~\mathrm{J/cm^2}$ (resp. $0.53~\mathrm{J/cm^2}$) for the uniform fiber (resp., for the Bragg fiber). Such values are within the range of demonstrated pump powers for subwavelength-scale waveguides fabricated in these materials \cite{Dorofeev2018NonlinearWriting, Ohishi2012FlatFiber, Liao2012DirectlyLength, DesAutels2008FemtosecondMaterials, Cardenas2015OpticalWaveguides, Tezuka2016Mid-infraredFiber} (we note that pump powers could be further reduced at the cost of smaller confinements, which may still be enough for long range guiding -- see Appendix~B). The resulting guiding energy is found to be $\hbar\Omega = 51.4~\mathrm{\mu eV}$ (resp., $325~\mathrm{\mu eV}$), with a fundamental guided wavefunction size of $\Delta r_e = 27.2 ~\mathrm{nm}$ (resp., $10.8~\mathrm{nm}$). In both cases, we therefore consider strong localization such that $a/\Delta r_e > 4.5$. 

Finally, we consider other limiting factors that constrain the maximal interaction length (see Fig.~\ref{fig:gqu-perspective}): For the uniform HCNF, $L_{\mathrm{int}}=4~\mathrm{cm}$, corresponding to a GVM length of $L_{\mathrm{GVM}}=3.6\mathrm{cm}$ (calculated from TE group velocity $v_{g,\mathrm{TE}} = 0.4416 c$ and electron velocity $v=0.7 c$), and propagation losses of $\approx0.5~\mathrm{dB/cm}$ (considering tellurite glass nanofibers \cite{Yang2006PhotonicGlasses}). For the Bragg fiber, the TE group velocity is matched with the electron velocity, so the only limiting factor is propagation losses, which are taken to be $\approx0.4-3~\mathrm{dB/cm}$ (SiC nanophotonic waveguides \cite{Yi2022SiliconPhotonics}). Hence, we choose to take $L_{\mathrm{int}}=1~\mathrm{cm}$ for this example. 

\section{Appendix D: Calculation of the free-electron--photon coupling}

In this Appendix, we provide details on the derivation of the free-electron--photon coupling ($g_Q$) in the context of this paper. In general, the coupling strength to any mode at wavenumber $q$ can be calculated via the Fourier integral over the z-component of its electric field as
\begin{equation}
\begin{split}
    \mathrm{g}_{q} &= \int d^2\boldsymbol{\rho}~\psi_f^*(\boldsymbol{\rho})\psi_i(\boldsymbol{\rho}) \\& \times \frac{e}{\hbar\omega(q)}\int_{0}^{L_{\mathrm{int}}}dz e^{-i\frac{\omega(q)}{v} z} E_z^{(s,q)}(\boldsymbol{\rho},z),
    \label{eq:gq}
\end{split}
\end{equation}
where the index $s$ labels the mode family and band number; $v$ is the electron velocity; and $\psi_{i/f}$ are the initial and final transverse electron wavefunctions. Eq.~(\ref{eq:gq}) is general, and can be converted to expressions commonly found in the literature \cite{Kfir2019EntanglementsRegime, BenHayun2021ShapingElectrons, Huang2023Electron-PhotonCircuits}. In most studies, it is common to assume that $\psi_f \approx\psi_i$ and further that $|\psi_i(\boldsymbol{\rho})|^2\approx\delta(\boldsymbol{\rho})$, treating the electron as a point-like particle in the transverse direction. Below we relax that assumption in order to account for higher-order transitions. 

The \textit{total} coupling $|g_{Q}|$ is often calculated over a discrete set of cavity modes $q_j$, for which $|g_{Q}|^2=\sum_j|\mathrm{g}_{q_j}|^2$ \cite{Feist2022Cavity-mediatedPairs} or even for a single mode $q_0$ \cite{Kfir2019EntanglementsRegime, BenHayun2021ShapingElectrons} for which $|g_{Q}|=|\mathrm{g}_{q_0}|$. However, unless stated otherwise, here we consider a continuous spectrum of waveguide modes, for which $|g_Q|^2=\int dq/2\pi |\mathrm{g}_q|^2$. The electric field complex amplitude is $\mathbf{E}^{(s,q)}(\mathbf{r})=\sqrt{\hbar\omega(q)/2\epsilon_0 A_{s,q}}e^{iqz}\mathbf{u}^{(s,q)}(\mathbf{r})$, where $\epsilon_0$ is the vacuum permittivity, $A_{s,q}$ the mode area, and $\mathbf{u}^{(s,q)}(\mathbf{r})$ the mode function, which is a Bloch function (along $z$) for the Bragg (periodic) fiber. For the Bragg fiber, we can further decompose the Bloch function on a Fourier basis as $\mathbf{u}^{(s,q)}(\mathbf{r})=\sum_m \mathbf{u}^{(s,q)}_m (\boldsymbol{\rho})e^{i2\pi m z/\Lambda}$, which simplifies to $\mathbf{u}^{(s,q)}(\mathbf{r})=\mathbf{u}^{(s,q)}_0(\boldsymbol{\rho})$ in the uniform case. In general, only one Fourier order $m$ of the mode's Bloch function contributes to the coupling of a given mode to the electron. Using Eq.~(\ref{eq:gq}) and taking the $dq/2\pi$ integral over all $|\mathrm{g}_q|^2$, we find
\begin{equation}
\begin{split}
    |g_{Q}|^2 &=\frac{\alpha}{\tilde{A}} \frac{L_{\mathrm{int}}}{\lambda}\Big|\int d^2\boldsymbol{\rho}\psi_f^*(\boldsymbol{\rho})\psi_i(\boldsymbol{\rho})u_{m,z}^{(s,q_0)}(\boldsymbol{\rho})\Big|^2 \\& \times  \int_{-\infty}^{\infty} \frac{dx}{\pi} \mathrm{sinc}^2 \left[\left(1 -\frac{v_g}{v}\right)x-\frac{\omega''}{L_{\mathrm{int}}v}x^2 \right].
    \label{eq:gQu}
\end{split}
\end{equation}
In calculating Eq.~(\ref{eq:gQu}), we have assumed that physical quantities such as the mode function and mode area are slowly-varying in $q$, taking them outside the integral, and evaluating them for a mode at the phase-matching point having a free-space wavelength $\lambda$. Furthermore, $\alpha\approx 1/137$ is the fine structure constant, and $\tilde{A}=A/\lambda^2$ is the normalized mode area. 

The argument of the sinc function, corresponding to the phase-matching term, was obtained by Taylor expanding $\omega(q)$ around the phase-matching point, up to second order, where $v_g$ is the photonic group velocity and $\omega''$ denotes the second derivative of $\omega(q)$ evaluated at the phase-matching point (this quantity relates to the GVD parameter via $\omega''=-v_g^3 \beta_2$). The integral in Eq.~(\ref{eq:gQu}) takes a closed-form solution for the two types of phase-matching points we consider here: the intersection point (with $v\neq v_g$, neglecting the dispersion $\omega''\to 0$) and the tangency point ($v=v_g$ and $\omega''\neq 0$). Taking the integral in each of these respective limits recovers Eq.~(\ref{eq:gQu_maintext}) of the main text.

The parameters obtained for the TM modes in each of our examples are the following: for the intersection (resp., tangency) phase-matching point, we have electron velocity $v=0.7 c$ (resp., $v = 0.2575c$), TM group velocity $v_{g,\mathrm{TM}} = 0.4124 c$ (resp., $v_{g,\mathrm{TM}}$ matched to $v$), TM wavelength $\lambda_{\mathrm{TM}} = 646.53 \mathrm{nm}$ (resp., $\lambda_{\mathrm{TM}} = 423 \mathrm{nm}$), normalized TM mode area $\tilde{A} = 0.5175$ (resp., $\tilde{A} = 0.3775$), TM dispersion $\omega'' \approx 0 $ (resp., $\omega'' = 87.1 \mathrm{m^2/sec}$), overlap integral with the fundamental electron mode $\psi_{00}$ of $|\int d^2\boldsymbol{\rho}|\psi_{00}(\boldsymbol{\rho})|^2u_{0,z}^{(\mathrm{TM})}(\boldsymbol{\rho})|=0.3487$  (resp., $0.0154$ with the $m=2$ Fourier order), recoil at phase-matching of $q_0 =1.39\times 10^7$ (resp., $q_0 + 4\pi/\Lambda = 5.77\times 10^7 \mathrm{m^{-1}} $), corresponding Kerr nonlinearity of $\kappa =2\pi\times 1.77 \mathrm{GHz}$ (resp., $\kappa = 2\pi \times 30.06~\mathrm{GHz})$, and corresponding nonlinear phase $\delta_\text{NL} = 2\kappa (L_{\mathrm{int}}/v) = 1.35 \pi$ (resp., $\delta_\text{NL} = 15.88 \pi$).

\section{Appendix E : Modal analysis of HCNF}
In this Appendix, we detail the analytical solution of the hollow core nanofiber dispersion and guided mode distribution. We adapt (and generalize) the methods from, e.g., Refs. \cite{Ito1995OpticalFiber, Gao2019BoundGratings, GovindP.Agrawal2021FIBER-OPTICSYSTEMS}. If the permittivity of the structure varies along $z$, $\epsilon=\epsilon(z)$, it is known that the full fields can be expressed in terms of their $z$-components \cite{Gao2019BoundGratings}, which satisfy the following equations:
\begin{align}
\begin{split}
&\nabla^2H_z +\epsilon(z)k_0^2 H_z =0 \\
&\nabla^2E_z +\epsilon(z)k_0^2 E_z + \frac{\partial}{\partial z}\left(\frac{\epsilon'}{\epsilon} E_z\right) = 0,
\end{split}
\end{align}    
where $k_0 = \omega/c$ and prime denotes differentiation with respect to $z$. If the dielectric region is periodic with period $\Lambda$, the solutions to the above equations can be expanded in terms of Bessel and Hankel functions of order $l$ and angular momentum $e^{il\phi}$, multiplied by either plane waves $e^{i (q+2\pi m/\Lambda)z}$ (in the vacuum regions) or Bloch waves $e^{iqz}u_{\mathrm{TE},q}^{(m)}(z)$ and $e^{iqz}u_{\mathrm{TM},q}^{(m)}(z)$ (in the periodic dielectric region):
\begin{align}
\begin{split}
&E_z = e^{iqz}e^{il\phi}\\ &\times \sum_m \begin{cases} A_m \frac{J_l(\zeta_m\rho)}{J_l(\zeta_m a)}e^{i(2\pi m/\Lambda)z} &\rho\leq a\\
    \left[B_m \frac{H_l^{(1)}(\gamma_m\rho)}{H_l^{(1)}(\gamma_m a)} + C_m \frac{H_l^{2)}(\gamma_m\rho)}{H_l^{(2)}(\gamma_m b)}\right]u_{\mathrm{TM},q}^{(m)}(z) &a \leq \rho \leq b \\ 
    D_m \frac{H_l^{(1)}(\zeta_m\rho)}{H_l^{(1)}(\zeta_m b)}e^{i(2\pi m/\Lambda)z} &\rho \geq b \end{cases} \\
&H_z = \frac{1}{\mu_0 c} e^{iqz}e^{il\phi}\\ &\times \sum_m \begin{cases} E_m \frac{J_l(\zeta_m\rho)}{J_l(\zeta_m a)}e^{i(2\pi m/\Lambda)z} &\rho\leq a\\
    \left[F_m \frac{H_l^{(1)}(\gamma_m\rho)}{H_l^{(1)}(\gamma_m a)} + G_m \frac{H_l^{2)}(\gamma_m\rho)}{H_l^{(2)}(\gamma_m b)}\right]u_{\mathrm{TE},q}^{(m)}(z) &a \leq \rho \leq b \\ 
    H_m \frac{H_l^{(1)}(\zeta_m\rho)}{H_l^{(1)}(\zeta_m b)}e^{i(2\pi m/\Lambda)z} &\rho \geq b \end{cases},
\end{split}
\end{align}  
where $\zeta_m = \sqrt{k_0^2-(q+2\pi m/\Lambda)^2}$, and $\eta_m,~\gamma_m$ are eigenvalues for the equations satisfied by the Bloch functions \cite{Gao2019BoundGratings}:
\begin{align}
\begin{split}
&\left[(iq+\partial_z)^2+\epsilon(z)k_0^2\right]u_{\mathrm{TE},q}^{(m)}=\eta_m^2 u_{\mathrm{TE},q}^{(m)} \\
&\left[(iq+\partial_z)^2+\epsilon(z)k_0^2 + (iq + \partial_z) \frac{\epsilon'}{\epsilon}\right]u_{\mathrm{TM},q}^{(m)}=\gamma_m^2 u_{\mathrm{TM},q}^{(m)}
\end{split}
\end{align}  
The second equation is not self-adjoint, but can be transformed into one if we instead consider modified TM Bloch functions $v_{\mathrm{TM},q}^{(m)}=\sqrt{\epsilon(z)}u_{\mathrm{TM},q}^{(m)}$ which satisfy:
\begin{equation}
    \left[(iq+\partial_z)^2+\epsilon(z)k_0^2 -\sqrt{\epsilon(z)} \left(\frac{1}{\sqrt{\epsilon}}\right)''\right]v_{\mathrm{TM},q}^{(m)}=\gamma_m^2 v_{\mathrm{TM},q}^{(m)}.
\end{equation}
One finds the other field components $E_{\phi},E_{\rho},H_{\phi},H_{\rho}$ from $E_z,H_z$ by using Maxwell's equations:
\begin{align}
\begin{split}
&\nabla\times \mathbf{E}=i\omega\mu_0 \mathbf{H} \\
&\nabla\times \mathbf{H}=-i\omega\epsilon_0 \epsilon(z) \mathbf{E}.
\end{split}
\end{align}  
We then impose a set of boundary conditions on the fields $E_z, H_{\phi}, H_z, E_{\phi}$ at $\rho=a,~b$. 

From this set of boundary conditions, a matrix $\mathbf{M}$ of size $8(2N+1)$-by-$8(2N+1)$ (where $N=0,1,2...$ is the cutoff number of harmonics used in the calculation) is constructed, acting on the vector $\mathbf{c}$ of the $8\times (2N+1)$ unknown coefficients $A_m, B_m, ..., H_m$ ($-N\leq m\leq N$), such that $\mathbf{M}\mathbf{c}=0$. We then numerically solve for the roots $q,\omega(q)$ of the determinant
\begin{equation}
    \det \mathbf{M}=0
\end{equation}
and find the corresponding zero-eigenvector $\mathbf{c}$ of $\mathbf{M}$ that yield the modal functions. 

The different mode families are characterized by different OAM numbers $l$: the TE and TM mode families have $l=0$, and they occupy separable subspaces having either $E_z =0$ (with $A_m,...,D_m =0$) or $H_z = 0$ (with $E_m,...,H_m =0$), respectively. The HE and EH mode families have $l\neq 0$, with both $E_z,~H_z$ nonzero. A choice of $l$ in our solution can help us ``isolate'' different mode families.

For better convergence, the abrupt change in $\epsilon(z)$ was replaced with a smooth one using a super-Gaussian function:
\begin{equation}
    \epsilon(z) = n_1^2 +(n_2^2-n_1^2)\exp\lbrace-[(z-\Lambda/2)/\sigma]^p/2\rbrace
\end{equation}
with $p=10$, where the duty cycle is $D = 1-2\sigma/\Lambda$. After the effective refractive indices $n_{\mathrm{eff},1},~n_{\mathrm{eff},2}$ are found for $n_1$ and $n_2$, the period and duty cycle are calculated by requiring the quarter-wavelength condition for the desired bandgap wavelength:
\begin{equation}
    n_{\mathrm{eff},1}(1-D)\Lambda = n_{\mathrm{eff},2}D\Lambda = \lambda_{\mathrm{BG}}/4
\end{equation}
Last, we compute the mode area according to:
\begin{equation}
    A=\frac{\int d^2 \mathbf{r}_T ~\epsilon(\mathbf{r}_T, z) |E(\mathbf{r}_T,z)|^2 }{\max {\epsilon(\mathbf{r}_T, z) |E(\mathbf{r}_T,z)|^2}}
\end{equation}
where, for a uniform HCNF $A$ is constant in $z$, and for a Bragg HCNF we average $A(z)$ over a unit cell.

\section{Appendix F : Derivation of Kerr-like Hamiltonian in free-electron--photon interactions}
In this Appendix, we derive the Hamiltonian describing free-electron--photon interaction. In the case of an electron trapped along a guiding structure (an ``electron fiber''), the system is effectively one-dimensional. In the non-relativistic limit, the Hamiltonian is given by:
\begin{equation}
H = \frac{p^{2}}{2m} + \sum_{k}^{}{\hslash\omega_{k}a_{k}^{\dagger}a_{k} + \sum_{k}^{}{\hslash\Omega_{k}(a_{k}b^{\dagger}_k + a_{k}^{\dagger}b_k),}}
\end{equation}
where \(\Omega_{k} = ev_{0}\sqrt{\hslash/2\epsilon_{0}\omega_{k}V}\) is the coupling constant for each mode, and $b_k = e^{-ikz}$ is the electron momentum ladder operator, with $[b_k,b_{k'}^{\dagger}]=0$~(Ref.~\cite{Kfir2019EntanglementsRegime}). When momentum is conserved, we may make a considerable simplification of the Hamiltonian. Consider the total momentum operator of the electron and field, denoted $P$:
\begin{equation}
    P = p + \sum_{k}^{}{\hslash ka_{k}^{\dagger}a_{k}}.
\end{equation}
It can be easily shown that this operator is conserved, i.e., ~\(\dot{P} = 0.\) Therefore, we may write
\begin{equation}
 p = P - \sum_{k}^{}{\hslash ka_{k}^{\dagger}a_{k}},   
\end{equation}
where \(P\) is the operator in the Schrödinger picture (equal to that in the Heisenberg picture at time zero) and write:
\begin{align}
    H = &\frac{\left( P - \sum_{k}^{}{\hslash ka_{k}^{\dagger}a_{k}} \right)^{2}}{2m} + \sum_{k}\hslash\omega_{k}a_{k}^{\dagger}a_{k} \nonumber \\ &+ \sum_{k}\hslash\Omega_{k}(a_{k}b_k^{\dagger} + a_{k}^{\dagger}b_k).
\end{align}

We note that there are many initial conditions (e.g., an electron momentum state
and any Fock state of the photons) for which the initial state is an eigenstate of \(P\). Any state can be decomposed into eigenstates of this operator. For an eigenstate of this operator, we have that
\(P = P(0) = \langle\psi(0)|P|\psi(0)\rangle\), which is a \emph{c­-number}. This considerably simplifies the Hamiltonian. First, let us expand the Hamiltonian as:

\begin{align}
 H_{P} = &\frac{P^{2}(0)}{2m} + \sum_{k}^{}\hslash\left( \omega_{k} - \frac{kP}{m} +\frac{\hbar k^2}{2m}\right)a_{k}^{\dagger}a_{k} + \nonumber \\ &\sum_{k}^{}{\hslash\Omega_{k}(a_{k} b_k^{\dagger} + a_{k}^{\dagger}b_k)} + \frac{\hslash^{2}}{2m}\sum_{k,q}^{}{kqa_{k}^{\dagger}a_{q}^{\dagger}a_{q}}a_{k}.   
\end{align}
In introducing the subscript, we have made it more explicit that the
Hamiltonian is parameterized by \(P = P(0)\). Next, let us introduce the
operator 
\begin{equation}
    A_{k}  = a_{k}b_{k}^{\dagger}.
\end{equation}
Using the commutation relations of \(b_k\), we find that {[}\(A_{k},A_{k'}^{\dagger}\rbrack = \delta_{k,k'}\):
these new operators have the same commutation relations as photon
operators. Using the relation
\(a_{k}^{\dagger}a_{k} = A_{k}^{\dagger}A_{k}\), therefore, we can write
the Hamiltonian as:
\begin{align}
H = &\frac{P^{2}(0)}{2m} + \sum_{k}^{}\hslash\left( \omega_{k} - \frac{kP}{m} +\frac{\hbar k^2}{2m}\right)A_{k}^{\dagger}A_{k} \nonumber \\ &+ \sum_{k}^{}{\hslash\Omega_{k}(A_{k} + A_{k}^{\dagger})} + \frac{\hslash^{2}}{2m}\sum_{k,q}^{}{kqA_{k}^{\dagger}A_{q}^{\dagger}A_{q}A_{k}}.
\label{eq:hamiltonian-Ak}
\end{align}
This reformulated Hamiltonian will form the basis for our non-perturbative multiphoton theory nonlinear electron-light interaction. Specifically, in our examples we focus on spontaneous emission, where the initial state is an electron with momentum $p(0)=mv$ and field in the vacuum state $\ket{0}$, which is an eigenstate of $P$ with $P(0)=p(0)=mv$. Substituting this into Eq.~(\ref{eq:hamiltonian-Ak}), while approximating $k\approx q \approx k_0$ in the nonlinear term (with $k_0$ denoting the momentum recoil felt by the electron at the phase-matching point), and eliminating the constant energy term $P(0)^2/2m$, we get an effective Kerr Hamiltonian:
\begin{align}
H_{\mathrm{eff}} = \sum_{k}\hbar \Delta_k A_{k}^{\dagger}A_{k}  + \sum_{k}^{}{\hslash\Omega_{k}(A_{k} + A_{k}^{\dagger})} + \hbar\kappa N(N-1),    
\end{align}
where $\kappa = \hbar k_0^2 / 2m$ is the effective Kerr nonlinearity,
\begin{align}
\Delta_k=\omega_{k} - kv +\frac{\hbar k^2}{2m}
\end{align}
denotes the free-electron--photon detuning and
\begin{align}
N = \sum_{k}^{}{A_{k}^{\dagger}A_{k}}
\end{align}
is the total excitaion number. Expanding the detuning around the phase-matching point $k_0,\omega_0$ up to second order, we find
\begin{equation}
\begin{split}
\Delta_k & \approx (v_g - v +\frac{\hbar}{m} k_0)(k-k_0)+\frac{1}{2}\left(\frac{\hbar}{m} + \omega''\right)(k-k_0)^2 \\ & \approx (v_g - v )(k-k_0)+\frac{1}{2}\omega''(k-k_0)^2,
\end{split}    
\end{equation}
where $\omega''=(d^2\omega/dk^2)_{k=k_0}$. As in Appendix D, we can set $v\neq v_g$ and $\omega''\to 0$ for an intersection point, and $v= v_g$ with $\omega'' \neq 0$ for a tangency point.

This Hamiltonian can readily deal with initial states which are not eigenstates of \(P\). Let's look at the example of a Gaussian electron wavepacket $c_p$, centered around some $p_0$ with momentum uncertainty $\Delta p$, coupled to the vacuum state. In this case, our initial wavefunction of the free-electron--photon system, is given by
\begin{equation}
  \left| \psi \right\rangle = \sum_{p}^{}{c_{p}\left| p \right\rangle\ \left| 0 \right\rangle.}  
\end{equation}
The evolution of a state \(\left| p \right\rangle|0\rangle\) is given by the above Hamiltonian, setting \(P = p\). Thus, formally one solves \(i\hslash\partial_{t}\left| \psi \right\rangle = H_{P}|\psi\rangle\) for each momentum state. The evolution of the state $\ket{\psi}$ is given by a linear superposition of these individual time evolutions weighted by $c_p$. 

We note, however, that the dynamics could be well-approximated by a Hamiltonian of a single subspace $P$, when certain assumptions are made on the initial electron wavefunction.

First, an electron momentum uncertainty $\Delta p$ can, in general, cause broadening of the phase-matching bandwidth. This introduces an additive detuning such that $\Delta_k\to \Delta_k +k p' /m$ (where $p'$ can be of the order of $\Delta p$), and accordingly an additional detuning phase-mismatch of the order of $(k\Delta p /m)\times (L/2v)$ (independent of the type of phase-matching point). If this additional detuning is small enough compared to the nonlinearity and if the additional phase mismatch is much smaller than $\pi$, we can take $P(0)\approx p_0$ to a very good approximation, integrate the Schrödinger equation and convolve the resulting state with the initial electron wavefunction (such an approach was used to fit theory with experimental data to a very high level of accuracy \cite{Dahan2021ImprintingElectrons}). Explicitly, in terms of electron energy uncertainty $\Delta E$, this condition reads
\begin{equation}
    \frac{\Delta E}{E}\ll \min{\left(\frac{\hbar \omega}{2E}, 2\beta \frac{\lambda}{L_{\mathrm{int}}}\right)}
    \label{eq:condition1}
\end{equation}
where $\lambda = 2\pi c/\omega$ is the free space wavelength and $\beta = v/c$. 

Second, we must also assume that we operate in the so-called "particle-like" regime of free-electron--light interaction \cite{Huang2023Electron-PhotonCircuits}, where the electron's coherent momentum uncertainty $\Delta p$ is much larger than the phase-matching bandwidth of the considered supermodes. This ensures that individual electron momenta do not get highly-entangled with individual photon momenta \textit{within} a given supermode, as would otherwise occur for plane-waves \cite{Karnieli2021TheParticle}. In terms of energy uncertainty, this condition reads as 
\begin{equation}
    \frac{\Delta E}{E} \gg   2\frac{\lambda_{\mathrm{dB}}}{L_{\mathrm{int}}} \begin{cases} 1/|1 -v_g/v| &\text{intersection}\\
    \sqrt{L_{\mathrm{int}}v/\pi \omega''} &\text{tangency}
    \end{cases}       
    \label{eq:condition2}
\end{equation}
where $\lambda_{\mathrm{dB}}=h/mv$ is the de Broglie wavelength of the electron. The resulting free-electron--photon entanglement is quasi-discrete (as opposed to a continuous entanglement between free-electron--photon momentum components). Such entanglement manifests as correlations between an optical excitation (e.g., a supermode centered around $\omega_0$) with $n$ photons, and a corresponding electron energy loss peak around $E -n\hbar\omega_0$ \cite{Huang2023Electron-PhotonCircuits}.  

In the following, we shall assume both conditions Eq.~(\ref{eq:condition1}) and Eq.~(\ref{eq:condition2}) hold, which can be readily achieved in our examples for $\Delta E \leq 0.1 \mathrm{eV}$ (and even more so with experimentally-available electron monochromators \cite{Krivanek2019ProgressEELS,Yannai2023LosslessRadiation} allowing energy uncertainties down to tens of meV). The aforementioned assumptions allow us to simulate the dynamics through Hamiltonian Eq.~(\ref{eq:hamiltonian-Ak}), and, if needed, trace out the electron degrees of freedom by taking the trace over the discrete electron energy loss peaks.

\section{Appendix G : Supermode analysis of free-electron--photon nonlinear dynamics}
In this Appendix, we derive a supermode theory of our nonlinear dynamics. Supermodes are typically used to describe quantum nonlinear dynamics in multimode systems, such as synchronously-pumped optical parametric oscillator~\cite{DeValcarcel2006MultimodeCombs}. We write the Heisenberg equations for $A_k$ as:
\begin{equation}
    \dot{A}_k = -i\Delta_k A_k -2i\kappa N A_k -i\Omega_k^*.
\end{equation}
In this equation, the nonlinear term is proportional to the total excitation number and is therefore invariant upon unitary transformation on the mode basis. We can thus choose a supermode basis, represented by wavepacket operators
\begin{equation}
    w_n (t) = \int dk \tilde{w}_n (k) e^{i\Delta_k t}A_k (t),
    \label{eq:dynamics-wavepacket}
\end{equation}
where the $\tilde{w}_n (k)$'s are an orthonormal, complete basis of wavepackets, such that $\sum_n w_n^\dagger w_n = \int dk A_k^\dagger A_k = N$. The supermodes satisfy the following dynamics
\begin{equation}
    \dot{w}_n  = -2i\kappa w_n -is_n (t)   
\end{equation}
with driving terms given by 
\begin{equation}
    s_n(t) = \int dk \Omega_k^* e^{i\Delta_k t} \tilde{w}_n (k).
    \label{eq:driving-terms}
\end{equation}

We first consider linear dynamics ($\kappa = 0$) over interaction time $T$ (with time $t\in [-T/2,T/2]$). At $t=T/2$, the field populates a $single$ supermode, denoted as $w_0$, with wavepacket
\begin{equation}
    \tilde{w}_0(k) = \frac{\Omega_k T}{|g_Q|} \mathrm{sinc}{\frac{\Delta_k T}{2}}
\end{equation}
such that $w_n (T/2) = w_n (-T/2) -i|g_Q|\delta_{n,0}$, recovering a known result of the quantum theory of the photon-induced near field electron microscopy effect, which uses instead the scattering matrix formalism \cite{Kfir2019EntanglementsRegime, BenHayun2021ShapingElectrons}. 

In the presence of the nonlinear term $\kappa\neq0$, the dynamics may be described by multiple supermodes or approach that of a single supermode, depending on the type of phase-matching point considered. In any case, in the following we shall consider an orthonormal basis of wavepackets containing $\tilde{w}_0(k)$, wherein most of the energy should still be contained.  

The general procedure for finding the supermode basis $\lbrace\tilde{w}_n\rbrace$ is the following: choose an orthonormal basis $\lbrace\tilde{\psi}_n\rbrace$, containing a basis vector $\tilde{\psi}_0$ which maximally overlaps with $\tilde{w}_0$ (or that's identical to it, in which case $\lbrace\tilde{w}_n\rbrace\equiv\lbrace\tilde{\psi}_n\rbrace$). If necessary, one can construct $\lbrace\tilde{w}_n\rbrace$ from $\lbrace\tilde{w}_0\rbrace$ and $\lbrace\tilde{\psi}_n\rbrace/\lbrace\tilde{\psi}_0\rbrace$ using a Gram-Schmidt process. Then, calculate an auxiliary set of driving terms for the $\lbrace\tilde{\psi}_n\rbrace$ basis, given by
\begin{equation}
\bar{s}_n(t) = \int dk \Omega_k^* e^{i\Delta_k t} \tilde{\psi}_n (k)   
\end{equation}
from which the sources for the dynamics of $w_n$ of Eq.~(\ref{eq:dynamics-wavepacket}) could be readily inferred via
\begin{equation}
    s_n(t) = \sum_m (\tilde{\psi}_m|\tilde{w}_n) \bar{s}_m(t),
\end{equation}
where $(\cdot|\cdot)$ denotes inner product between complex functions. 

We now outline the different examples, and consider $\Omega_k = \Omega$ uniform, for simplicity. Whenever needed, we use the auxiliary Hermite function basis
\begin{equation}
    \tilde{\psi}_n(k) = \frac{1}{\sqrt{2^nn!\sqrt{\pi}\sigma}}e^{-k^2/2\sigma^2}H_n\left(\frac{k}{\sigma}\right),
\end{equation}
find the scale $\sigma$ for which the overlap between $\tilde{\psi}_0(k)$ and $\tilde{w}_0(k)$ is maximal, and proceed according to the procedure above.

\textit{Phase-matching to a mode continuum.} First, consider phase-matching to a continuum of modes in a waveguide. For an intersection point, with linear dispersion $\Delta_k = \delta v k$ (where $\delta v = v_e -v_g$), we find 
\begin{equation}
    \bar{s}_n(t) = \frac{|g_Q|}{T}  \frac{i^n\sqrt{2\sigma'}}{\sqrt{2^n n!\sqrt{\pi}}} \exp{\left[-\frac{1}{2}\left(\sigma'\frac{2t}{T}\right)^2\right]}H_n\left(\sigma' \frac{2t}{T}\right)
\end{equation}
where $\sigma' = \sigma \delta v T/2 \approx 1.4$ is the optimal scaling for the intersection point. In deriving $\bar{s}_n(t)$ we used the identity
\begin{equation}
\int_{-\infty}^{\infty} dx e^{ixt} e^{-x^2/2}H_n(x)=\sqrt{2\pi} i^n e^{-t^2/2}H_n (t).
\end{equation}

For a tangency point, with quadratic dispersion $\Delta_k = \omega'' k^2/2$, we find  
\begin{equation}
\begin{split}
\bar{s}_{2n}(t) &= \frac{|g_Q|}{T} \frac{(2n)!}{\sqrt{2^{2n}(2n)!}n!} \sqrt{\frac{3}{2}\sigma'}  \\ & \times\frac{\exp\left[i(2n+\frac{1}{2})\arctan(2\sigma'^2\frac{2t}{T})\right]}{\left[1+(2\sigma'^2\frac{2t}{T})^2\right]^{\frac{1}{4}}},    
\end{split}
\end{equation}
where $\bar{s}_{2n+1}(t) = 0$, and this time $\sigma' = \sigma \sqrt{\omega''T}/2=\sqrt{\sqrt{3}/2}\approx0.93$ is the optimal scaling. In deriving $\bar{s}_n(t)$ we used the identity 
\begin{equation}
\int_{-\infty}^{\infty} dx e^{-cx^2}H_{2n}(x)=\sqrt{\frac{\pi}{c}} \frac{(2n)!}{n!} \left(\frac{1-c}{c}\right)^n,~~\mathrm{Re} (c)>0.
\end{equation}

\textit{Phase-matching to a discrete set of cavity modes.} Now, consider the phase matching to a discrete set of cavity modes $k_j = (\pi/L)j$, where $L$ is the cavity length. In that case, the dispersion is 
\begin{equation}
    \Delta_j = \delta v \frac{\pi}{L} j = \frac{\pi v_e}{L} \left(1-\frac{v_g}{v_e}\right)j \equiv \Delta j,
\end{equation}
and the fundamental supermode is
\begin{equation}
    \tilde{w}_0(j) = \frac{\Omega T}{|g_Q|} \mathrm{sinc} \left[\left(1-\frac{v_g}{v_e} \right)\frac{\pi}{2}j\right],
\end{equation}
where we assumed that the electron traverses the full length of the cavity such that $T=L/v_e$, and introduced the effective free spectral range (FSR)
\begin{equation}
    \label{effFSR}
    \Delta = \frac{\pi v_e}{L}\left(1-\frac{v_g}{v_e}\right).
\end{equation}
approaching the true FSR for $|v_e| \ll |v_g|$. Next, we consider the special case where 
\begin{equation}
    1-\frac{v_g}{v_e} = 2m,
\end{equation}
for an integer $m$ other than 0, for which $\Delta = 2\pi m/T$ and
\begin{equation}
    \tilde{w}_0(j) = \frac{\Omega T}{|g_Q|} \delta_{j,0}.
\end{equation}
In that case, our supermode basis $\lbrace \tilde{w}_n (j)\rbrace$ coincides with the cavity mode basis $\lbrace {\delta_{j,n}} \rbrace$ - as it is already a complete orthonormal basis containing $\tilde{w}_0$, and we can readily calculate the driving terms as 
\begin{equation}
    s_n(t) = \frac{|g_Q|}{T}\exp\left({in \frac{2\pi m}{T} t}\right) 
\end{equation}
Finally, the nonlinear cavity dynamics can take two limiting forms: when the nonnlinear detuning $2\kappa$ is an integer multiple of $\Delta$, after each recoil the electron becomes phase-matched to a neighbouring cavity mode, and thus cascaded phase-matching to different cavity modes can occur. However, if the detuning $2\kappa$ is a half-integer multiple of $\Delta$, then after a single recoil, the electron is maximally-detuned from neighbouring modes:
\begin{equation}
    2\kappa = \begin{cases} n\Delta & \mathrm{cascade} \\
    (n+1/2)\Delta & \mathrm{maximal~detuning}
    \end{cases}  
\end{equation}
where in the maximally-detuned case, the limits of $m\gg 1$ (or $v_e \ll v_g$) and $\delta_{\mathrm{NL}} = 2\kappa T \gg 2\pi$ reduce to the single-mode Jaynes-Cummings interaction for slow electrons \cite{Karnieli2023Jaynes-CummingsPhotons}.  Both cases are illustrated in Fig.~\ref{fig:tangency} of the main text.

\section{Funding} A.K. is supported by the VATAT-Quantum fellowship by the Israel Council for Higher Education; the Urbanek-Chodorow postdoctoral fellowship by the Department of Applied Physics at Stanford University; the Zuckerman STEM leadership postdoctoral program; and the Viterbi fellowship by the Technion. C.~R.-C. is supported by a Stanford Science Fellowship. N.R. acknowledges the support of a Junior Fellowship from the Harvard Society of Fellows. S. F. acknowledges the support from the Department of Energy. (Grant No. DE-FG02-07ER46426).  
\section{Acknowledgments} The authors acknowledge fruitful discussions with Ido Kaminer and Zhexin Zhao.
\section{Disclosures} The authors declare no conflict of interest.
\section{Data availability} Data may be obtained from the authors upon reasonable request. 
\bibliography{references}
\bibliographystyle{ieeetr}

\end{document}